%% file: alicepreprint_CDS.tex
\documentclass[ALICE,manyauthors]{cernphprep}
\usepackage[comma,square,numbers,sort&compress]{natbib}
\usepackage{hyperref}
\usepackage{lineno}
\usepackage{epsfig} 
\usepackage{subfigure}
\usepackage{verbatim}
\usepackage{color}
\usepackage{soul}
\usepackage{caption}  
\usepackage{array}    
\usepackage{multirow} 
\usepackage[T1]{fontenc}
\usepackage{amsbsy}
\usepackage{graphicx}
\usepackage{here}

\input{aliases.tex}

\begin{document}%

\begin{titlepage}
\PHyear{2019}
\PHnumber{070}      
\PHdate{09 April}  

%

\title{Measurement of charged jet cross section in pp collisions at $\pmb{\sqrt{s}=5.02}\ \mathbf{TeV}$}
\ShortTitle{Charged jets in pp collisions at $\sqrts=5.02~\tev$}   

\Collaboration{ALICE Collaboration\thanks{See Appendix~\ref{app:collab} for the list of collaboration members}}
\ShortAuthor{ALICE Collaboration} 

\begin{abstract}
  \input{abstract.tex}
\end{abstract}
\end{titlepage}
\setcounter{page}{2}

%
%
\input{paper.tex}               
%
%

\newenvironment{acknowledgement}{\relax}{\relax}
\begin{acknowledgement}
\section*{Acknowledgements}
\input{fa_2019-02-11.tex}    
\end{acknowledgement}

\newpage

\bibliographystyle{utphys}   
\bibliography{alicepreprint_CDS}

\newpage
\appendix

\section{The ALICE Collaboration}
\label{app:collab}
\input{2019-02-11-Alice_Authorlist_2019-Feb-11.tex}  

\newpage
\section{Appendix}
\label{app:appendix}
\input{appendix.tex}

\end{document}

%% file: aliases.tex
\newcommand{\sqrts}{\sqrt{s}}

\newcommand{\tev}{\mathrm{TeV}}

%% file: abstract.tex
The cross section of jets reconstructed from charged particles is measured in the transverse momentum range of $5<p_\mathrm{T}<100\ \mathrm{GeV}/c$ in pp collisions at the center-of-mass energy of $\sqrt{s} = 5.02\ \mathrm{TeV}$ with the ALICE detector. The jets are reconstructed using the anti-$k_\mathrm{T}$ algorithm with resolution parameters $R=0.2$, $0.3$, $0.4$, and $0.6$ in the pseudorapidity range $|\eta|< 0.9-R$. The charged jet cross sections are compared with the leading order (LO) and to next-to-leading order (NLO) perturbative Quantum ChromoDynamics (pQCD) calculations. It was found that the NLO calculations agree better with the measurements.  The cross section ratios for different resolution parameters were also measured. These ratios increase from low $p_\mathrm{T}$ to high $p_\mathrm{T}$ and saturate at high $p_\mathrm{T}$, indicating that jet collimation is larger at high $p_\mathrm{T}$ than at low $p_\mathrm{T}$. These results provide a precision test of pQCD predictions and serve as a baseline for the measurement in Pb$-$Pb collisions at the same energy to quantify the effects of the hot and dense medium created in heavy-ion collisions at the LHC.

%% file: paper.tex
\section{Introduction}
\label{sec:intro}
\input{introduction.tex}

\section{Experimental setup and data sample}
\label{sec:detector}
\input{detector.tex}



\section{Monte Carlo simulation}
\label{sec:mc}
\input{mc.tex}

\section{Inclusive charged jet cross section}
\label{sec:jetrec}
\input{jetrec.tex}

\section{Unfolding of detector effects}
\label{sec:unfolding}
\input{unfolding.tex}
\section{Systematic uncertainties}
\label{sec:Syst}
\input{systematics.tex}

\section{Results}
\label{sec:results}
\input{results.tex}

%

\section{Conclusion}
\label{sec:conclusion}
\input{conclusion.tex}


%% file: introduction.tex
In quantum chromodynamics (QCD), jets are defined as collimated showers of particles resulting from the fragmentation of hard (high-momentum transfer $Q$) partons (quarks and gluons) produced in short-distance scattering processes. Jet cross section measurements provide valuable information about the strong coupling constant, $\alpha_{\mathrm s}$, and the structure of the proton~\cite{Giele,Warburton}. In addition, inclusive jet production represents a background to many other processes at hadron colliders. Therefore, the predictive power of fixed-order perturbative QCD (pQCD) calculations of jet production is relevant in many studies in high-energy collisions, and the inclusive jet cross section measurements in proton-proton collisions provide a clean test of pQCD. Jet production in $\mathrm{e}^{+}\mathrm{e}^{-}$, $\mathrm{ep}$, $\mathrm{p\overline{p}}$, and pp collisions is quantitatively described by pQCD calculations~\cite{Verbytskyi:2018ulf,Kogler:2011np,Bhatti:2010bf}. 

Jets also constitute an important probe for the study of the hot and dense QCD matter created in high-energy collisions of heavy nuclei. In nucleus-nucleus (A$-$A) collisions, high-$p_{\rm T}$ partons penetrate the colored medium and lose energy via induced gluon radiation and elastic scattering (see~\cite{JHEP03_2013-080} and references therein) while in proton-nucleus (p$-$A) collisions, jet production may be modified by cold nuclear matter (CNM) effects~\cite{PRC92_2015_054911,EPJC74_2014_2951,PLB749_2015_68,PLB776_2018_249}. 
Furthermore, in high multiplicity pp and p$-$A collisions, jet production could be modified even if the collision system is small. The measurements of inclusive jets in pp collisions thus provide a baseline for similar measurements in A$-$A, p$-$A and high multiplicity pp and p$-$A collisions.

The measured jet properties are typically well reproduced by many general-purpose Monte Carlo (MC) event generators~\cite{Sjostrand:2016bif}. The unprecedented beam energy achieved at the Large Hadron Collider (LHC)~\cite{Evans:2008zzb} in pp collisions enables an extension of the energy range of jet production cross section and property measurements carried out at lower energies. Such measurements enable further tests of QCD and help in tuning of MC event generators. Inclusive jet production cross sections have been measured in collisions of hadrons at the $\mathrm{Sp\bar{p}S}$ and Tevatron colliders at various center-of-mass energies. The latest and most precise results at $\sqrt{s} = 1.96\ \mathrm{TeV}$ are detailed in Refs.~\cite{PRD78_2008_052006, PRD85_2012_052006}. At the LHC at CERN, the ALICE, ATLAS and CMS
collaborations have measured inclusive jet cross sections in proton$-$proton collisions at center-of-mass energies of $\sqrt{s} =2.76\ \mathrm{TeV}$~\cite{PLB722_2013, EPJC73_2013_2509, EPJC76_2016_265} , $7\ \mathrm{TeV}$ \cite{EPJC71_2011_1512, PRD87_2013_112002} and $8~\mathrm{TeV}$~\cite{1706.03192, JHEP03_2017_156}. Recently, the ATLAS and CMS collaborations have measured the inclusive jet cross sections at $\sqrt{s} =13\ \mathrm{TeV}$~\cite{EPJC76_2016_451, 1711.02692}. 

This paper presents the measurements of the inclusive charged jet cross sections in proton$-$proton collisions at a center-of-mass energy of $\sqrt{s} =5.02\ \mathrm{TeV}$ by the ALICE Collaboration at the LHC. 
The inclusive charged jet cross sections are measured double-differentially as a function of the jet transverse momentum, $p_\mathrm{T}$, and absolute jet pseudorapidity, $|\eta|$. The jets are reconstructed using the anti-$k_\mathrm{T}$ jet clustering algorithm~\cite{JHEP04_2008_063} with resolution parameter values of $R = 0.2$, $0.3$, $0.4$, and $0.6$. The inclusive charged jet cross sections are measured in the kinematic region of $5 < p_\mathrm{T} < 100\ \mathrm{GeV}/c$ and pseudorapidity of $|\eta|< 0.9-R$. 
The analysis is restricted to jets reconstructed solely from charged particles, hereafter called charged jets. 
Charged particles with momenta down to $p_\mathrm{T} > 0.15\ \mathrm{GeV}/c$ are used in the jet reconstruction of different $R$ values, thereby allowing us to test perturbative and non-perturbative aspects of jet production and fragmentation as implemented in MC event generators \cite{Dasgupta:2007wa,Chatrchyan:2014gia}. 
ALICE reported similar measurements of charged jet production in pp~\cite{ALICE:2014dla,Acharya:2018eat}, p$-$Pb~\cite{Adam:2015hoa, Adam:2016jfp}, and Pb$-$Pb collisions \cite{JHEP03_2014_013} using data from the first LHC run. 

A brief description of the ALICE detector and the selected data sample are introduced in Sec.~\ref{sec:detector}. MC simulations and theoretical calculations used for comparison to data are presented in Sec.~\ref{sec:mc}. The cross section definition is given in section~\ref{sec:jetrec} and the unfolding procedure is described in Sec.~\ref{sec:unfolding}. Systematic uncertainties on the cross section measurements are addressed in Sec.~\ref{sec:Syst}. Finally, the results without and with underlying event (UE) subtraction are presented and discussed in Sec.~\ref{sec:results} and Appendix~\ref{app:appendix}, respectively.

%% file: detector.tex
ALICE (A Large Ion Collider Experiment) is a dedicated heavy-ion experiment at the LHC, CERN. A detailed description of the detectors can be found in ~\cite{ALICEexp}. The detector components used in the data analysis presented in this publication are outlined here.

The ALICE detector comprises a central barrel (pseudorapidity $|\eta|<0.9$ coverage over full-azimuth) immersed in a uniform $0.5\ \mathrm{T}$ magnetic field along the beam axis ($z$) supplied by the large solenoid magnet. The forward-rapidity plastic scintillator counters are positioned on each side of the interaction point, covering pseudorapidity ranges $2.8<\eta<5.1$ (V0A) and $-3.7<\eta<-1.7$ (V0C), and they are used for determination of the interaction trigger. The central barrel contains a set of tracking detectors: a six-layer high-resolution silicon Inner Tracking System (ITS) surrounding the beam pipe (from inside outward: the Silicon Pixel (SPD), Drift (SDD), and Strip (SSD) Detectors), and a large-volume ($5\ \mathrm{m}$ length, $5.6\ \mathrm{m}$ diameter) Time-Projection Chamber (TPC). 

The ITS and TPC space-points are combined to reconstruct tracks from charged particles over a wide transverse momentum range ($0.15\ <p_\mathrm{T}<100\ \mathrm{GeV}/c$). The selected tracks are required to have at least 70 TPC space-points out of a maximum of 159 possible and more than $60\%$ of the findable TPC space-points based on the track parameters. For the best momentum resolution, at least 3 track hits are required to be located in the ITS. The primary vertex position is reconstructed from charged particle tracks as described in~\cite{PrimVertex}. Only tracks originating from the primary vertex, called primary tracks, are used for jet reconstruction. These tracks are selected based on their distance of closest approach to the primary vertex of the interaction (smaller than $2.4\ \mathrm{cm}$ and $3.2\ \mathrm{cm}$ in the transverse plane and along the beam axis, respectively).

To fully compensate the loss of tracking efficiency with the SPD dead areas and recover good momentum resolution,
tracks without any hit in either of the two SPD layers, referred as "hybrid tracks",  
are also retained but constrained to the primary vertex~\cite{BelikovVERTEX2018}. The tracking efficiency estimated from a full simulation of the detector (see Sec.~\ref{sec:mc}) is $80\%$ for $p_\mathrm{T}>0.4\ \mathrm{GeV/c}$, decreasing to $60\%$ at $0.15\ \mathrm{GeV/c}$. The momentum resolution is better than $3\%$ for hybrid tracks below $1\ \mathrm{GeV}/c$, and increases linearly up to $10\%$ at $p_\mathrm{T}=100\ \mathrm{GeV}/c$. 
 
The measurement presented here uses data from pp collisions at a center-of-mass energy of $\sqrt{s}=5.02~\,\mathrm{TeV}$ collected in 2015. During this period, minimum-bias (MB) events are selected using the high purity V0-based MB trigger 
~\cite{ALICEperf} which required a charged particle signal coincidence in the V0A and V0C arrays. The corresponding visible pp cross section was measured with the van der Meer technique to be $51.2\pm 1.2\ \mathrm{mb}$ \cite{ALICEvdm}. During the intensity ramp up, the instantaneous luminosity delivered by the LHC was successively leveled to $2\times 10^{29}\ \mathrm{cm^{-2}\,s^{-1}}$ and $10^{30}\ \mathrm{cm^{-2}\,s^{-1}}$ resulting in interaction rates of $10\ \mathrm{kHz}$ and $50\ \mathrm{kHz}$, respectively \cite{Muratori:2014ija}. The track quality was checked and it was found to be independent of interaction rates.

Further selection of MB events for offline analysis is made by
requiring a primary vertex position within $\pm 10\ \mathrm{cm}$ around the nominal interaction point to ensure full geometrical acceptance in the ITS for $|\eta|<0.9$. Pile-up interactions are maintained at an average number of pp interactions per bunch crossing below $0.06$ through beam separation in the horizontal plane. Residual pile-up events are rejected based on a multiple vertex finding algorithm using SPD information \cite{BelikovVERTEX2018}. After event selection, a data set of $103\times 10^6$ minimum-bias pp collisions corresponding to an integrated luminosity $\mathcal{L}_\mathrm{int}=2\ \mathrm{nb}^{-1}$ is used.

%% file: mc.tex
\label{secMC}

Monte Carlo (MC) event generators are used both for predictions of jet production to compare with data, and to provide simulations of detector performance for particle detection and reconstruction used to correct the measured distributions for instrumental effects. 
For the latter case, primary simulated events are generated with the PYTHIA8\cite{Pythia8} (PYTHIA 8.125, Monash 2013 tune~\cite{Pythia8Monash}) MC generator. Then particles are transported through the simulated detector apparatus with GEANT~3.21 \cite{A32_RefGeant3}. The simulated and real data are analyzed with the same reconstruction algorithms. 

The MC generators HERWIG~\cite{Herwig,A36_Herwig_1} (HERWIG~6.510) and  PYTHIA6 (PYTHIA~6.425 and several UE tunes defined as everything accompanying an event but the hard scattering)~\cite{Pythia} are used for variations of the detector response and systematic investigations 
of the MC correction factors as well as jet fragmentation and hadronization patterns (as described in section~\ref{sec:Syst}). For comparison with data, MC simulated samples with different tunes from PYTHIA6, PYTHIA8, and POWHEG merged with PYTHIA8 for the parton shower and hadronization \cite{powheg0,powheg1,powheg2,powheg3} are used. 


PYTHIA and HERWIG are both event generators based on leading order (LO) pQCD calculations of matrix elements for $2\to 2$ reactions of parton-level hard scattering. However, each generator utilizes different approaches to describe the parton shower and hadronization processes. HERWIG makes angular ordering a direct part of the evolution process and thereby takes coherence effects into account in the emission of soft gluons. PYTHIA6 is based on transverse-momentum-ordered showers \cite{A37_pt_ordered_showers} in which angular ordering is imposed by an additional veto. In PYTHIA6 the initial-state evolution and multiple parton-parton interactions are interleaved into one common decreasing $p_\mathrm{T}$ sequence. In PYTHIA8 the final-state evolution is also interleaved with initial-state radiation and multiparton interactions. Hadronization in PYTHIA proceeds via string breaking as described by the Lund model \cite{A38_Lund_model}, whereas HERWIG uses cluster fragmentation \cite{Kupco:1998fx}. 

The PYTHIA Perugia tune variations, beginning with the central tune Perugia-0 \cite{A31_PerugiaTunes}, are based on LEP, Tevatron, and SPS data. 
The PYTHIA6 Perugia 2011 family of tunes \cite{A31_PerugiaTunes}
belongs to the first generation of tunes that use LHC pp data at $\sqrt{s}=0.9$ and $7\ \mathrm{TeV}$.  For the PYTHIA8 Monash 2013 tune ~\cite{Pythia8Monash}, data at $\sqrt{s} = 8$ and $13\ \mathrm{TeV}$ are also used. The PYTHIA8 CUETP8M1 tune uses the parameters of the Monash Tune and fits to the UE measurements performed by CMS~\cite{CMSUE}. The HERWIG generator and PYTHIA6 tunes used in this work utilize the CTEQ5L parton distribution functions (PDFs) \cite{A40_CTEQ5L}. The PYTHIA8 Monash tune uses the NNPDF2.3 LO set \cite{PDF_NNPDF}. The uncertainty on the PDFs has been taken into account by the variation of the final results for the respective uncertainty sets of the PDFs.     

The POWHEG framework, an event-by-event MC, is used for next-to-leading order (NLO) pQCD calculations of $2\to 2$ and $2\to 3$ parton scattering at $\mathcal{O}(\alpha_{\rm s}^3)$. The outgoing partons from POWHEG are passed to PYTHIA8 event-by-event where the subsequent parton shower is performed. Double-counting of partonic configurations is inhibited by a matching scheme based on shower emission vetoing \cite{Buckley:2016bhy}. Contrary to fixed-order NLO calculations, the POWHEG MC approach has the advantage that the same selection criteria and jet finding algorithm can be used on the final state particle-level as used in the analysis of the real data; in particular, only charged particles can be selected. For the comparison with the measured differential jet cross sections, the CT14nlo PDF set is used~\cite{PDF_CT14NLO}. The dominant uncertainty in the parton-level calculation is given by the choice of renormalization, $\mu_\mathrm{R}$, and factorization scale, $\mu_\mathrm{F}$. The default value is chosen to be $\mu_\mathrm{R} = \mu_\mathrm{F} = p_\mathrm{T}$ of the underlying Born configuration, here a $2\to 2$ QCD scattering \cite{powheg0}. Independent variations by a factor of two around the central value are considered as the systematic uncertainty. For the POWHEG calculations, the PYTHIA8 A14 tune is used ~\cite{A14_CMS_PbPb_FF}.


%% file: jetrec.tex

\label{sec:jetReconstruction}
Jets are reconstructed from charged particles using the anti-$k_\mathrm{T}$ jet clustering algorithm \cite{A25_RefAntikt,A27_RefFastjet} with resolution parameters $R=0.2$, $0.3$, $0.4$, and $0.6$. The jet transverse momenta are calculated using a boost-invariant $p_{\rm T}$ recombination scheme as the scalar sum of their constituent transverse momenta. 
The bin-averaged differential inclusive charged jet cross section measured as a function of charged jet transverse momentum $p_{\rm T}^{\rm ch \; jet}$ in bins of pseudorapidity is defined as
\begin{equation}
  \frac{{\rm {d^{2}}}\sigma^{\rm ch \; jet}}{{\rm d} p_{\rm
      T}\rm{d}\eta} (p_{\rm T}^{\rm ch \; jet}) =
  \frac{1}{\mathcal{L}_{\rm int}}\frac{ \rm{d}{\it N}_{\rm
      jets}}{\rm {d} {\it p}_{\rm T} \rm{d} \eta} (p_{\rm
    T}^{\rm ch \; jet}), 
  \label{xsec-equation}
\end{equation}
where $\mathcal{L}_{\rm int}$ is the integrated luminosity given in section~\ref{sec:detector} and  $N_{\rm jets}$ is the number of jets reconstructed in bins of width $d p_{\mathrm T}$ in transverse momentum and $d \eta$ in pseudorapidity. One single bin of pseudorapidity $|\eta|< 0.9-R$ is considered in this analysis because of the limited coverage of the ALICE central barrel.  The measurements are performed in the kinematic range of $5\ <p_{\rm T}^{\rm ch \; jet}<100\ \mathrm{GeV}/c$.   

Jets observed in pp collisions are inevitably affected by the Underlying Event (UE) activity originating from Multiple Parton Interactions (MPI), fragmentation of beam remnants, and initial and final state radiation~\cite{Field:2012jv}. 
The UE can be characterized on an event-by-event basis by the amount of transverse momentum density $\rho_\mathrm{UE}$ in a `control region` cone of the same radius as the jet resolution parameter placed perpendicular to the leading jet axis, at the same pseudorapidity as the leading jet but offset by an azimuthal angle of $\pm \pi/2$ relative to the jet axis \cite{ALICE:2014dla}. To obtain the $\rho_\mathrm{UE}$, we calculate the sum of track $p_{\rm T}$ in a perpendicular cone which is defined with respect to a leading jet axis and divided by jet area as
\begin{equation}
        \rho_\mathrm{UE} = \sum_{i=0}^{n} p_{\rm T,i}^{\rm perp} / \pi R^{2}, 
  \label{rho_def}
\end{equation}

where $R$ is the jet resolution parameter and $p_{\rm T,i}^{\rm perp}$ is a transverse momentum of $i$th track in a perpendicular cone.

\begin{figure}[htbp]
 \begin{center}
  \includegraphics[width=130mm]{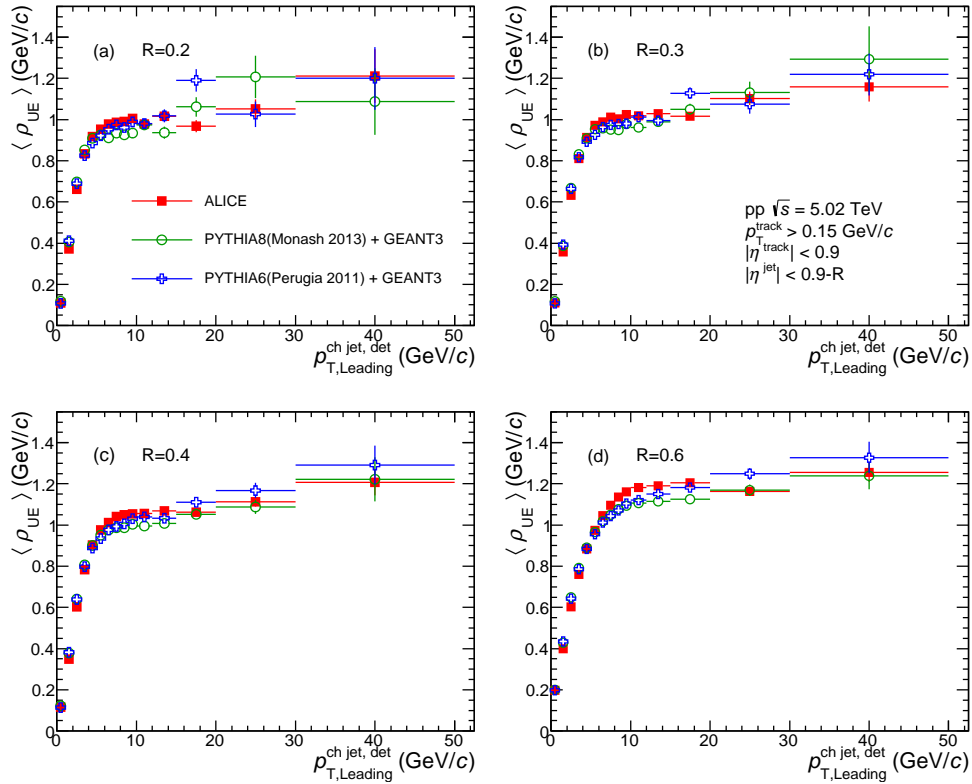}       
 \end{center}
 \caption{Dependence of the average $\rho_\mathrm{UE}$ on leading jet transverse momentum at detector level compared with predictions from PYTHIA (Perugia-2011 and Monash-2013 tunes) for the resolution parameter $R=0.2$ (a), $0.3$ (b), $0.4$ (c), and $0.6$ (d).}
 \label{Fig:UE}
\end{figure}

The average $\rho_\mathrm{UE}$ as a function of the event scale defined by the leading jet $p_\mathrm{T}$ is shown in Fig.~\ref{Fig:UE} for resolution parameters $R=0.2$, $0.3$, $0.4$, and $0.6$. 
The relative UE contribution increases with increasing jet transverse momentum. 
A steep rise of the UE activity in the transverse region is observed with increasing leading jet $p_\mathrm{T}$ followed by a slower rise above $10\ \mathrm{GeV}/c$ which suggests a weaker correlation with the hard process~\cite{JHEP07_2012_116}. The average UE also has a weak dependence on jet finding resolution parameters. 
While the asymptotic value of $\langle \rho_\mathrm{UE} \rangle$ is located close to $1\ \mathrm{GeV}/c$ for resolution parameter from $0.2$ up to $0.4$, it increases by $20\%$ for $R=0.6$, probably due to the contamination from jet regions which might arise for such a large cone size. Fig.~\ref{Fig:UE} compares the data to the recent tunes of the PYTHIA MC event generators as a function of detector level jet $p_{\rm T}$. The measured transverse momentum density can be reproduced by different PYTHIA tunes within 5\%, i.e. a slight underestimation from the Monash-2013 tune when approaching the slowly rising region. A similar observation was reported by an earlier publication of UE measurements using leading particles instead of jets~\cite{JHEP07_2012_116}.


All the observables studied in this paper are measured both with and without UE corrections, with the former presented in Appendix~\ref{app:appendix}, and the latter in the body of the paper. The impact of the UE subtraction on the inclusive jet spectrum can be seen in Fig.~\ref{Fig:UERatio}. A systematic uncertainty on the $\rho_\mathrm{UE}$ measurement was estimated to be $5\%$~\cite{ALICE:2014dla} resulting in a $2\%$ uncertainty on the UE subtracted jet cross section at $p_{\rm T}^{\rm ch \; jet}=5\ \mathrm{GeV}/c$ and decreasing for higher jet transverse momentum. Furthermore, as a reference for constructing jet nuclear modification factors in Pb$-$Pb collisions \cite{JHEP03_2014_013,Adam:2015ewa}, leading-track biased jet spectra are made available in Appendix in Fig.~\ref{Fig:XsecBiased}.

Finally, the differential inclusive charged jet cross sections are corrected for detector resolution and unfolded to the charged particle level (section~\ref{sec:unfolding}) to allow for a direct comparison to theoretical predictions (section~\ref{sec:results}).

%% file: unfolding.tex
The measurement of the steeply falling jet transverse momentum spectrum is affected 
by the imperfect track reconstruction efficiency and finite track momentum resolution of the detector. The inference of the true spectrum from the smeared one, a process usually called unfolding, requires construction of a detector response matrix. The jet production yields are corrected by the unfolding method~\cite{DAgostini:1994zf}, as implemented in the RooUnfold package~\cite{Adye:2011gm}. A 2-dimensional detector response matrix maps the transverse momentum of particle-level charged jets clustered from stable charged particles produced by a MC event generator ($p_\mathrm{T}^\mathrm{jet,particle}$) to the detector-level jets reconstructed from tracks after full GEANT3-based detector simulation ($p_\mathrm{T}^\mathrm{jet,detector}$). The entries of the response matrix are computed by matching particle- and detector-level jets geometrically, according to the distance $d = \sqrt{\Delta\eta^{2}+\Delta\phi^{2}}$ between the jet axes. The anti-$k_\mathrm{T}$ jet finding algorithm is used for both particle-level and detector-level jet reconstruction.

\begin{figure*}[htbp]
 \begin{center}
  \includegraphics[width=0.415\textwidth]{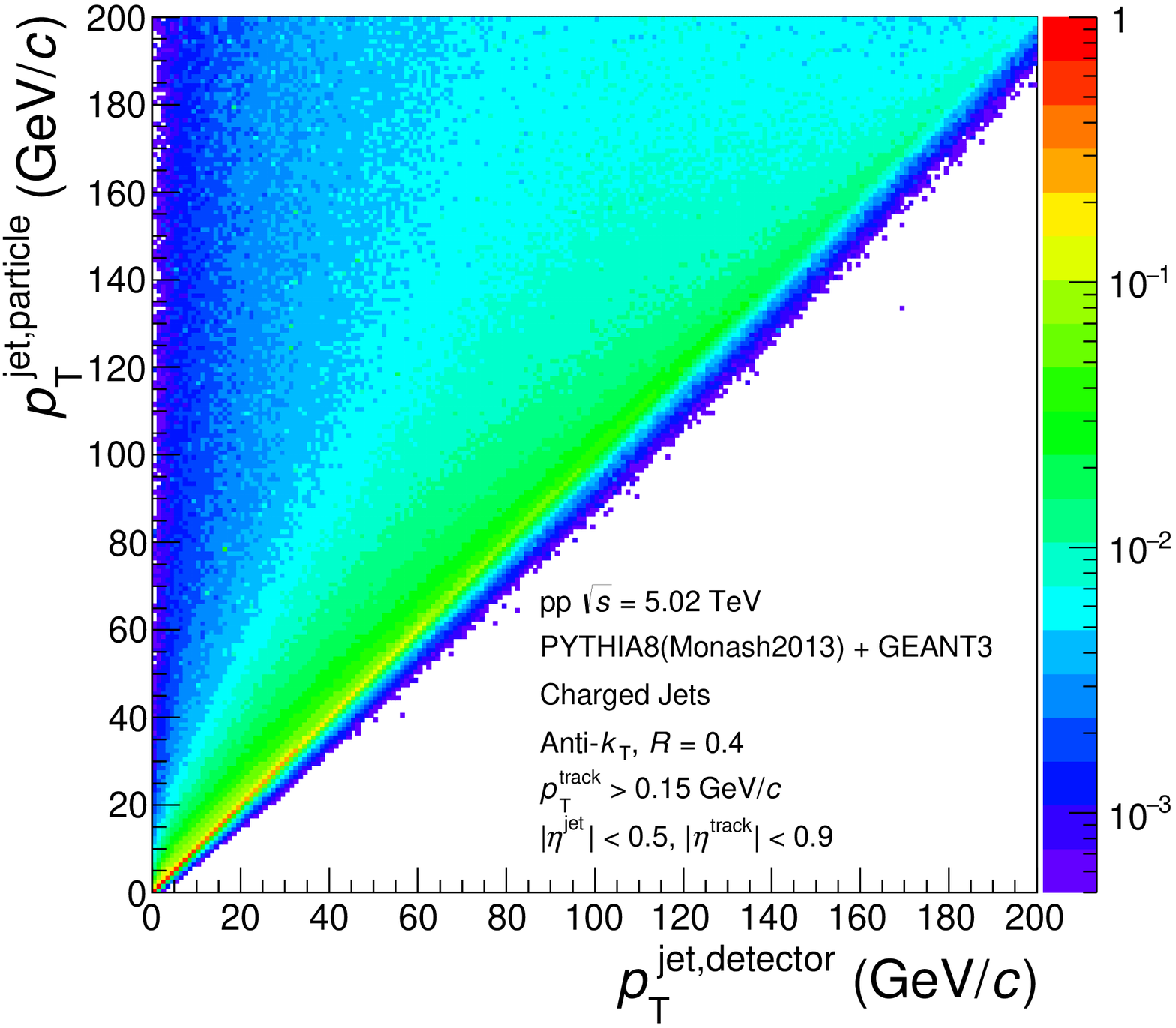}     
  \includegraphics[width=0.54\textwidth]{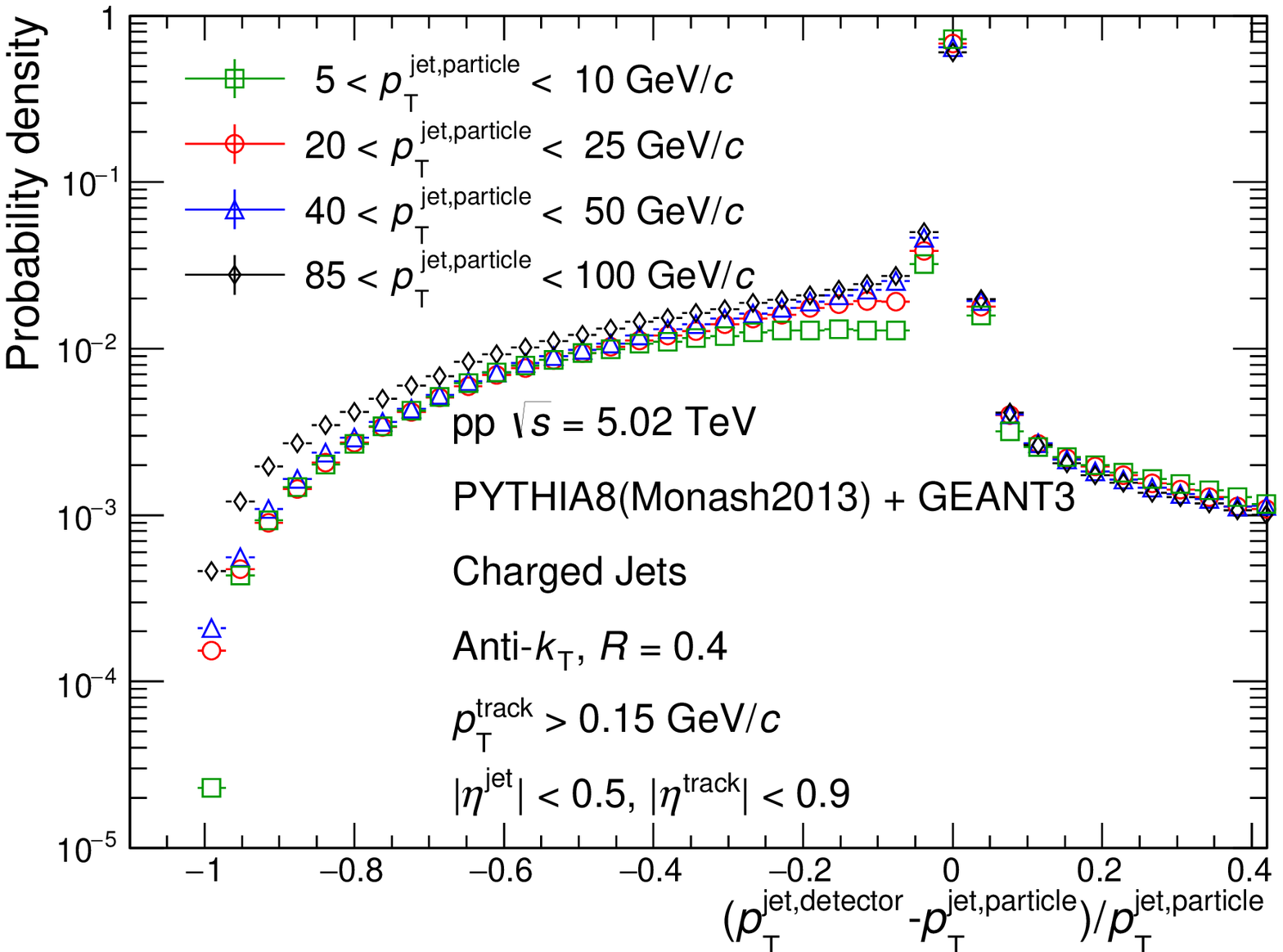}
 \end{center}
 \caption{Left: Detector response matrix for $R = 0.4$ charged jets. Right: 
Probability distribution of the relative momentum difference of simulated ALICE detector response to charged jets in pp collisions at $\sqrt{s} = $ 5.02 TeV for four different $p_\mathrm{T}$ intervals. Charged jets are simulated using PYTHIA8 Monash-2013 and reconstructed with the anti-$k_\mathrm{T}$ jet finding algorithm with $R = 0.4$.}
 \label{Fig:DetRespResidual}
\end{figure*}

The probability of reconstructing a charged jet at a given detector level $p_\mathrm{T}$ as a function of the particle level $p_\mathrm{T}$ is shown in Fig.~\ref{Fig:DetRespResidual} (left) for charged jets with $R = 0.4$, demonstrating the detector response matrix.
The probability distribution is derived from this detector response matrix and shown in Fig.~\ref{Fig:DetRespResidual} (right) for four different $p_\mathrm{T}^\mathrm{jet,particle}$ intervals. The distributions have a pronounced peak at zero ($p_\mathrm{T}^\mathrm{jet,detector} = p_\mathrm{T}^\mathrm{jet,particle}$). The tracking $p_\mathrm{T}$ resolution induces upward and downward fluctuations with equal probability, whereas the finite detection efficiency of the charged particles results in an asymmetric response. 

In this analysis, an unfolding approach relying on a Singular Value Decomposition (SVD) of the detector response matrix is used in order to reduce sizable statistical fluctuations that are introduced by instabilities in the inversion procedure~\cite{A44_unfold-svd}. This technique also produces a complete covariance matrix, along with its inverse, which allows for full uncertainty propagation. In addition, a Bayesian unfolding \cite{DAgostini:1994zf} was carried out for cross-check and systematic error assessments. Consistent results were obtained with both methods. 
To validate the unfolding process, and identify potential biases, closure tests are performed which compare the unfolded detector-level distribution to the particle-level truth in the MC simulation. Consistency of the unfolding procedure is also ensured by folding the solution to the detector level and comparing it to the uncorrected distribution used as input. No significant difference is found.

%% file: systematics.tex
The various sources of systematic uncertainties and their corresponding estimates obtained in this study are summarized in Table~\ref{tab:sysTable} and discussed in detail in the following sections. All systematic uncertainties listed in Table~\ref{tab:sysTable} are considered as uncorrelated except the unfolding one. Therefore, these systematic uncertainties were treated separately and their respective contributions are added in quadrature. In the ratio of the measured cross sections for different radii, the uncertainties from the same source cancel out partially, and the remaining relative difference is taken as the systematic uncertainty on the ratio. The total uncertainty on the jet cross section ratio is determined by adding the remaining contributions from different resources in quadrature.

\newcommand{\specialcell}[2][l]{\begin{tabular}[#1]{@{}l@{}}#2\end{tabular}}  
\renewcommand{\arraystretch}{1.5}

\begin{table}[bht]

\small 

\centering  

\begin{tabular}{ 
   m{1.45cm}             | 
   r                       
   m{1.25cm}<{\centering}   
   m{1.25cm}<{\centering}   
   m{1.25cm}<{\centering}   
   m{1.75cm}<{\centering}   
   m{1.75cm}<{\centering}   
   m{0.8cm}<{\centering} } 
 
 \hline 
 
Jet resoultion parameter &  \rule{0.2cm}{0cm} \specialcell[c]{  Jet $p_\mathrm{T}$ bin\\
   (GeV/$c$) } & Tracking efficiency (\%) & Track $p_{\rm T}$ resolution (\%)& Unfolding (\%) &  Normalization (\%) & Secondaries (\%)  & Total (\%) \\ 
\hline 

%
%
%
%
%
%

\multirow {4}{*}{   $R$ = 0.2   }
&  5$-$6   &  1   & negligible & 1.4 & 2.3  & 2.4    & 3.7      \\ \cline{2-8}
&  20$-$25 &  2.6 & negligible  & 2.3 & 2.3  & 2.2    & 4.7      \\ \cline{2-8}
&  40$-$50 &  5.2 & negligible  & 3.8 & 2.3  & 2.5    & 7.3      \\ \cline{2-8}
& 85$-$100 &  10  & negligible  & 7.8 & 2.3  & 2.6    & 13.1     \\ \cline{2-8} 

\hline 

\multirow {4}{*}{ $R$ = 0.3 } 
&  5$-$6   &  1.5 & 0.1  & 2.9 & 2.3  & 2.2     & 4.6      \\ \cline{2-8}
&  20$-$25 &  4.1 & 0.1  & 3.4 & 2.3  & 2.3     & 6.3      \\ \cline{2-8}
&  40$-$50 &  6.2 & 0.1  & 4.3 & 2.3  & 2.6     & 8.3      \\ \cline{2-8}
& 85$-$100 &  8.4 & 0.1  & 7.0   & 2.3  & 2.7   & 11.5     \\ \cline{2-8}  

\hline 

\multirow {4}{*}{ $R$ = 0.4   } 
&  5$-$6   &  0.9 & 1.9  & 1.9 & 2.3  & 2.1    & 4.2      \\ \cline{2-8}
&  20$-$25 &  3.7 & 1.9  & 1.8 & 2.3  & 2.4    & 5.6       \\ \cline{2-8}
&  40$-$50 &  5.4 & 1.9  & 2.5 & 2.3  & 2.5    & 7.2       \\ \cline{2-8}
& 85$-$100 &  7.5 & 1.9  & 4.5 & 2.3  & 2.8    & 9.6        \\ \cline{2-8} 
\hline

\multirow {4}{*}{ $R$ = 0.6   } 
&  5$-$6   &  3.4  & 1  & 2.1 & 2.3  & 1.9     & 5.1     \\ \cline{2-8}
&  20$-$25 &  5.7  & 1  & 1.7 & 2.3  & 2.6     & 6.9     \\ \cline{2-8}
&  40$-$50 &  6.8  & 1  & 2.2 & 2.3  & 2.6     & 8       \\ \cline{2-8}
& 85$-$100 &  8.3  & 1  & 4.0   & 2.3  & 2.7   & 9.9     \\ \cline{2-8} 

\hline 
\end{tabular}
\captionsetup{width = 1.0\textwidth}
\caption{\normalsize Summary of the systematic uncertainties for a selection of jet transverse momentum bins.  
}

\normalsize 

\label{tab:sysTable}
\end{table}

\subsection{Tracking efficiency and momentum resolution}
\label{sec:SysTrkEffRes}
To evaluate the impact of the limited track reconstruction efficiency and momentum resolution on the jet cross sections, a fast detector response simulation is used to reduce computing time. The efficiency and resolution are varied independently, and a new response matrix is computed for each variation. The detector-level distributions are then unfolded, and the resulting differences are used as systematic uncertainties. 
The relative systematic uncertainty on tracking efficiency is estimated to be 3\% based on the variations of track selection criteria. The track efficiency contributes a relative systematic uncertainty of up to 8\% on the jet cross sections since
it introduces a reduction and smearing of the jet momentum scale.

The systematic uncertainty of the jet cross sections due to the tracking efficiency uncertainty, which is the dominant source of uncertainty, increases with increasing jet $p_{\rm T}$ and resolution parameter, while the systematic uncertainty due to momentum resolution is neglibile with no $p_{\rm T}$ dependence and a weak dependence on the jet resolution parameter.  


\subsection{Unfolding}
\label{sec:SysUnfolding}
The reconstructed jet transverse momentum spectra presented in this paper are unfolded using a detector response computed with the Monash 2013 tune of the PYTHIA8 event generator \cite{Pythia8Monash}. This particular choice of MC event generator affects the detector response by influencing the correlation between the particle- and detector-level quantities used to evaluate the response matrix. Such a MC event generator dependence is quantified by comparing the unfolded spectrum using the default response matrix and generator prior with those obtained with the HERWIG and PYTHIA6 event generators with Perugia-0 and Perugia-2011 tune \cite{A31_PerugiaTunes}. This comparison is accomplished by using detector responses from fast simulation. The resulting uncertainty is on the order of $5\%$.

The SVD unfolding method~\cite{A44_unfold-svd}, which is the default approach used in this analysis, is regularized by the choice of an integer valued parameter which separates statistically significant and non-significant singular values of the orthogonalized response matrix. 
The regularized parameter is tuned for each cone radius parameter, separately. 
To estimate the related systematic uncertainty, the regularization parameter is varied by $\pm 2$ around the optimal value. The unfolded results are stable against regularization parameter variations with a maximum deviation of $1\%$ at high-$p_\mathrm{T}$.

Lastly, the SVD unfolded spectra are compared with the results obtained with the Bayesian unfolding method~\cite{DAgostini:1994zf}. 
Within uncertainties, the solutions of both unfolding methods are consistent. 

The uncertainties discussed above are added in quadrature and referred to as the unfolding systematic uncertainty in Table~\ref{tab:sysTable}.

\subsection{Cross section normalization}
\label{sec:SysNorm}
A systematic uncertainty on the integrated luminosity measurement of $2.3\%$ \cite{ALICEvdm} is propagated to the cross section as fully correlated across all $p_\mathrm{T}$ bins. Therefore, it cancels out in the ratio of cross sections.

\subsection{Contamination from secondary particles}
\label{sec:SysSecondary}
Contamination from secondary particles produced by weak decays of strange particles (e.g.\ $K^{\mathrm 0}_{\mathrm S}$ and $\Lambda$),
photon conversions, or hadronic interactions in the detector material, and decays of charged pions is significantly reduced by the requirement on the distance of closest approach of the tracks to the primary vertex point. The uncertainty due to the secondary contribution corresponds to a jet transverse momentum scale uncertainty of $0.5\%$~\cite{ALICE:2014dla,Acharya:2018eat}.



%% file: results.tex

\subsection{Charged jet cross sections}
\label{sec:spectra}
\begin{figure*}[htbp]
 \begin{center}
 \includegraphics[width=0.8\textwidth]{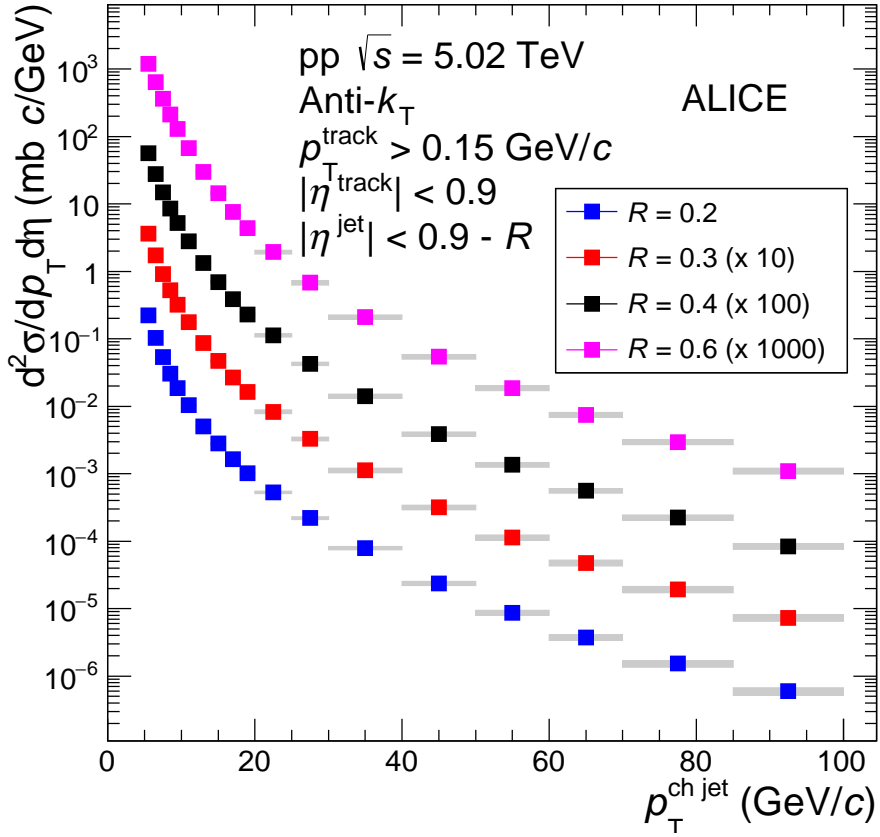}
 \end{center}
 \caption{Charged jet differential cross sections in pp collisions at $\sqrt{s}$ = 5.02 TeV after detector effect corrections. Statistical uncertainties are displayed as vertical error bars. The total systematic uncertainties are shown as shaded bands around the data points. Data are scaled to enhance visibility.}
 \label{Fig:Xsec}
\end{figure*}

The inclusive charged jet cross sections using the anti-$k_\mathrm{T}$ jet finding algorithm in pp collisions at $\sqrt{s} = 5.02\ \mathrm{TeV}$ are fully corrected for detector effects and are presented in Fig.~\ref{Fig:Xsec}. In this study, the inclusive charged jet cross sections are reported for jet resolution parameters $R = 0.2$, $0.3$, $0.4$, and $0.6$. The choice of $R$ is driven by which aspects of jet formation are investigated since the relative strength of perturbative and non-perturbative (hadronization and underlying event) effects on the jet transverse momentum distribution show a strong $R$-dependence \cite{Dasgupta:2007wa}. Pseudorapidity ranges are limited to $|\eta|< 0.9-R$ to avoid edge effects at the limit of the tracking detector acceptance.

\begin{figure*}[htbp]
 \begin{center}
  \includegraphics[width=0.85\textwidth]{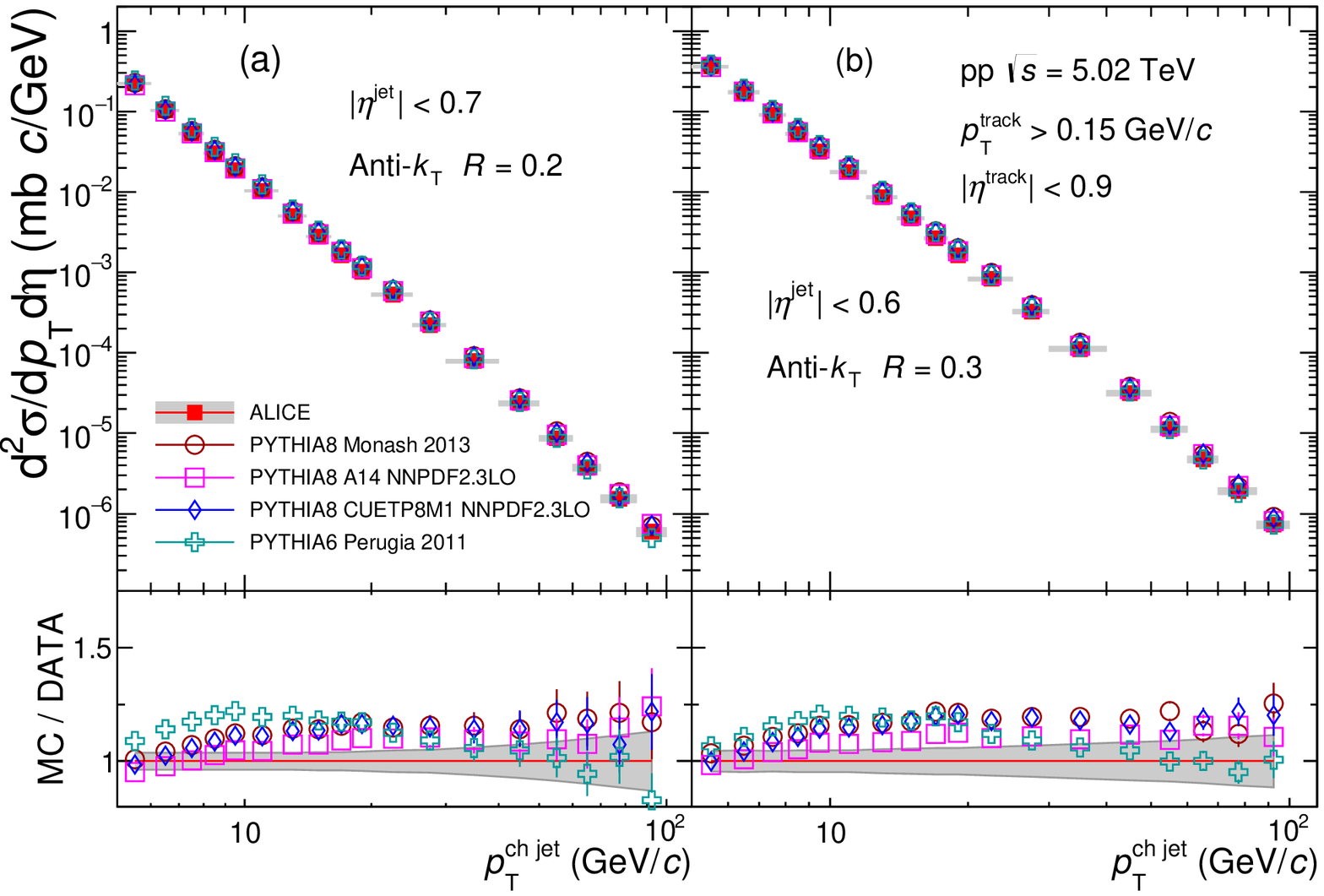} \\  
  \includegraphics[width=0.85\textwidth]{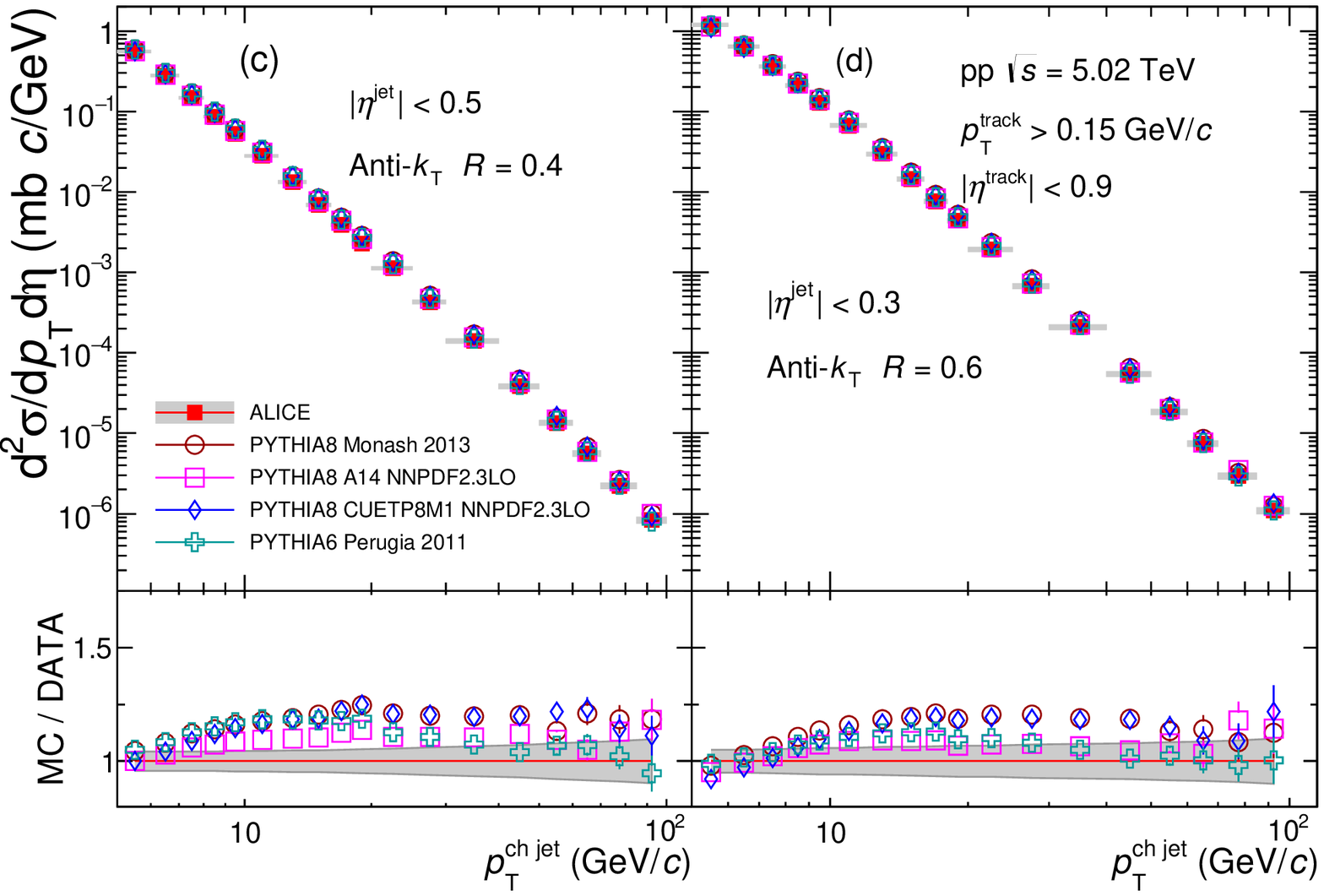}  
 \end{center}
 \caption{Comparison of the charged jet cross section to LO MC predictions with different 
jet resolution parameter $R=0.2$ (a), $0.3$ (b), $0.4$ (c), and $0.6$ (d). Statistical uncertainties are displayed as vertical error bars. The systematic uncertainty on the data is indicated by a shaded band drawn around unity. The red lines in the ratio correspond to unity.}
 \label{Fig:XsecCompMC}
\end{figure*}

\begin{figure*}[htbp]
 \begin{center}
  \includegraphics[width=0.85\textwidth]{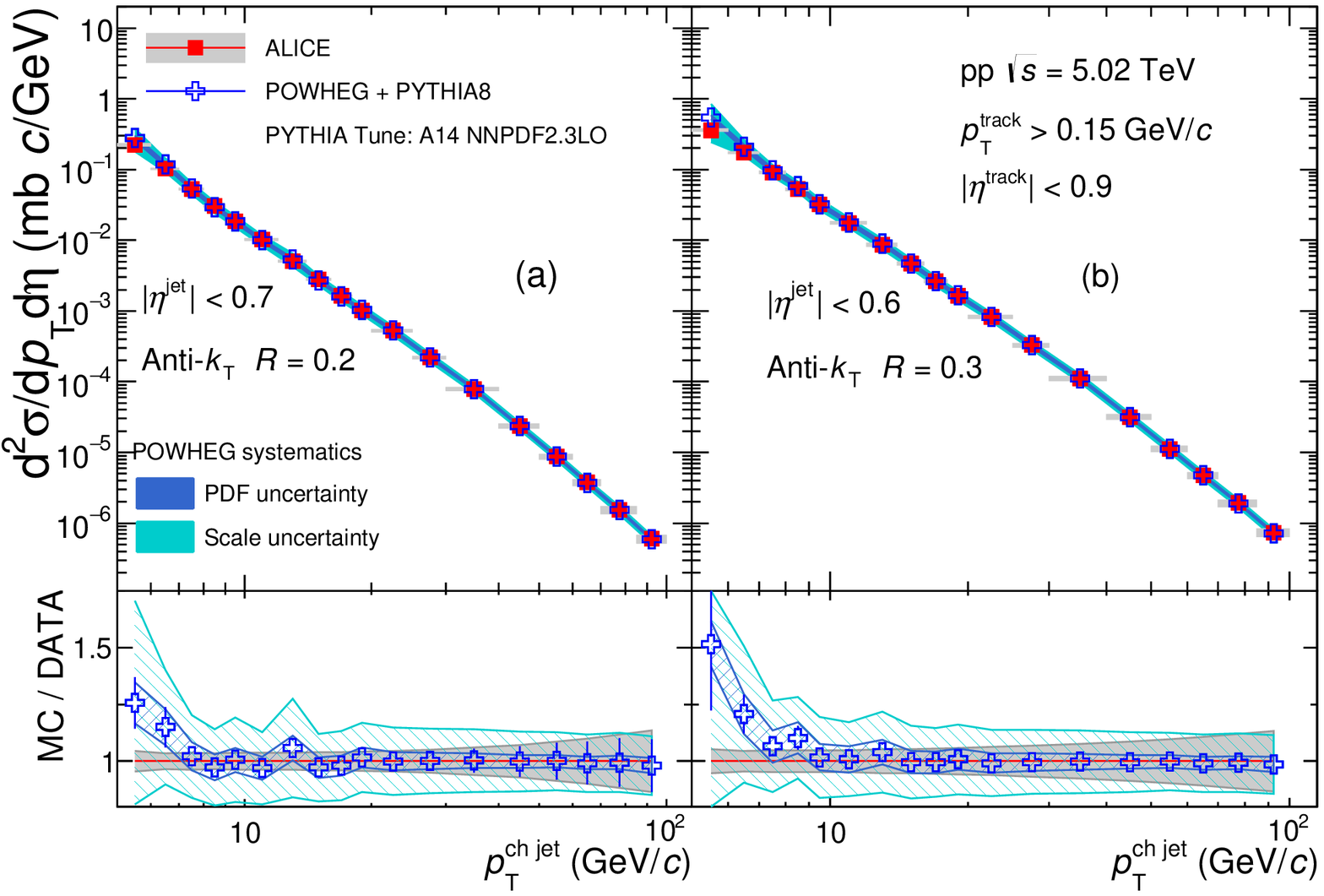} \\  
  \includegraphics[width=0.85\textwidth]{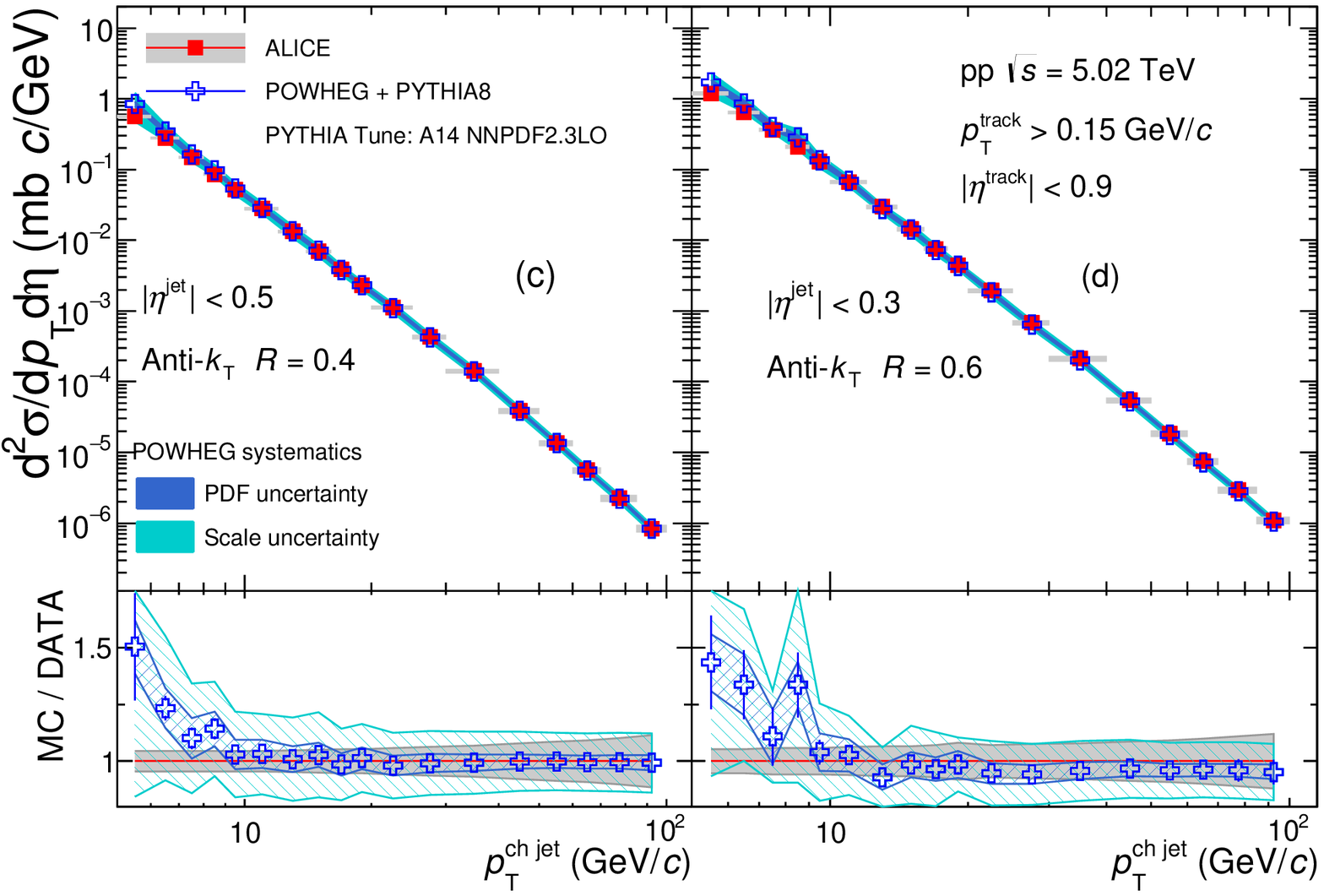}  
 \end{center}
 \caption{Comparison of the charged jet cross section to NLO MC predictions (POWHEG+PYTHIA8) with different 
jet resolution parameter $R=0.2$ (a), $0.3$ (b), $0.4$ (c), and $0.6$ (d). Statistical uncertainties are displayed as vertical error bars. The systematic uncertainty on the data is indicated by a shaded band drawn around unity. The red lines in the ratio correspond to unity.}
 \label{Fig:XsecCompMCNLO}
\end{figure*}

The differential cross sections of charged jets reconstructed using different jet resolution parameters $R$ are compared with LO PYTHIA predictions in Fig.~\ref{Fig:XsecCompMC}.
Fig.~\ref{Fig:XsecCompMCNLO} shows the comparison with POWHEG predictions. The ratios of the MC distributions to measured data are shown in the bottom panels. The model predictions qualitatively describe the measured cross sections, but fail to reproduce the shape over the entire jet transverse momentum range. The comparison between data and models is similar to earlier measurements at a lower center of mass energy~\cite{ALICE_chJets7TeV}. Although NLO corrections to inclusive single jet production improve the LO prediction and the NLO predictions agree within 10\% with the data in the studied phase space, the NLO prediction still disagrees with the data at the lowest kinematic phase space by up to 50\%, with very large theoretical uncertainty at low transverse momentum as shown in Fig.~\ref{Fig:XsecCompMCNLO}. At this low $p_\mathrm{T}$ region below 10 GeV/$c$, non-perturbative effects, such as soft particle production, multi-parton interactions, and fragmentation function bias play a role, which makes the comparison with theoretical models difficult.
Studies of next-to-next-to-leading order (NNLO) corrections using antenna subtraction~\cite{nnlojet} indicate that NNLO calculations should significantly reduce the systematic uncertainty from scale variations once they become available. 
Therefore, it is expected that a detailed theory-experiment comparison will be performed in the future using NNLO QCD corrections. This comparison will contribute to our understanding of pQCD processes.



\subsection{Ratio of charged jet cross sections}
\label{sec:Ratio}
\begin{figure*}[htbp]
 \begin{center}
   \includegraphics[width=0.8\textwidth]{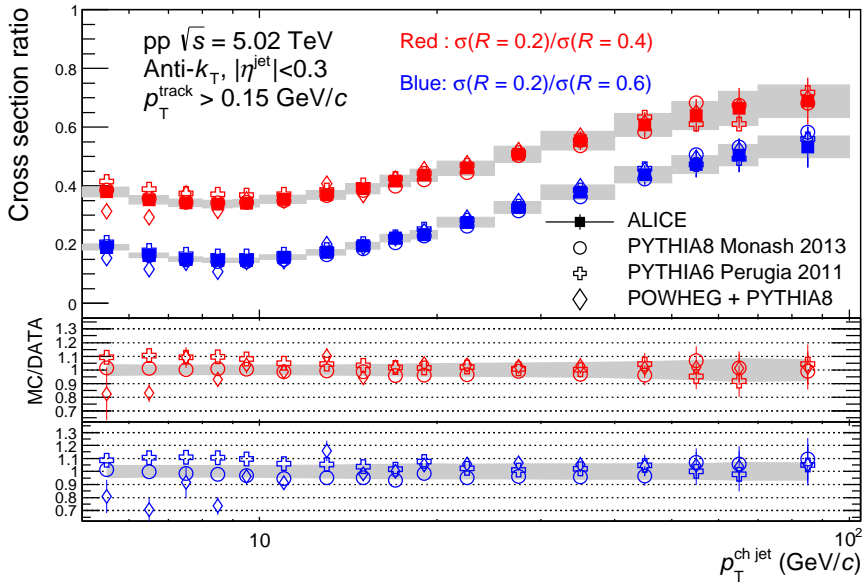}          
 \end{center}
 \caption{Charged jet cross section ratios for $\sigma(R=0.2)/\sigma(R=0.4)$ (Red) and $\sigma(R=0.2)/\sigma(R=0.6)$(Blue) in comparison with LO (PYTHIA) and NLO event generators with matched parton showers and modelling of hadronization and the UE (POWHEG+PYTHIA8). The systematic uncertainty of the cross section ratio is indicated by a shaded band drawn around data points. No uncertainties are drawn for theoretical predictions for better visibility.}
 \label{Fig:XsecRatio}
\end{figure*}

Figure~\ref{Fig:XsecRatio} shows the ratios of inclusive charged jet cross sections for jets reconstructed with a resolution parameter of $R = 0.2$ to those with $R = 0.4$ and 
$R = 0.6$. In order to compare the ratios within the same jet pseudorapidity range, the ratios are studied within $|\eta| < 0.3$, which coincides with the fiducial jet acceptance for the largest resolution parameter studied ($R=0.6$). Statistical correlations between the numerator and denominator are avoided by using exclusive subsets of the event sample. This observable relates directly to the relative difference between the jet $p_\mathrm{T}$ distributions when using two different resolution parameters and provides insights into the interplay between perturbative and non-perturbative effects. The departure from unity, which is due to the emission of QCD radiation, decreases as jet collimation increases at high transverse momentum. 
From the experimental point of view, the observable is less sensitive to experimental systematic uncertainties, 
and consequently the comparisons between theoretical predictions and data are less ambiguous for this observable than for inclusive spectra~\cite{PLB722_2013}.
The measured ratios are also compared with PYTHIA and POWHEG calculations in Fig.~\ref{Fig:XsecRatio}. Both models give a good description of the data within 10\%, stressing the significance of jet parton showers beyond higher-order matrix element calculations. 

\begin{figure*}[htbp]
 \begin{center}
   \includegraphics[width=0.49\textwidth]{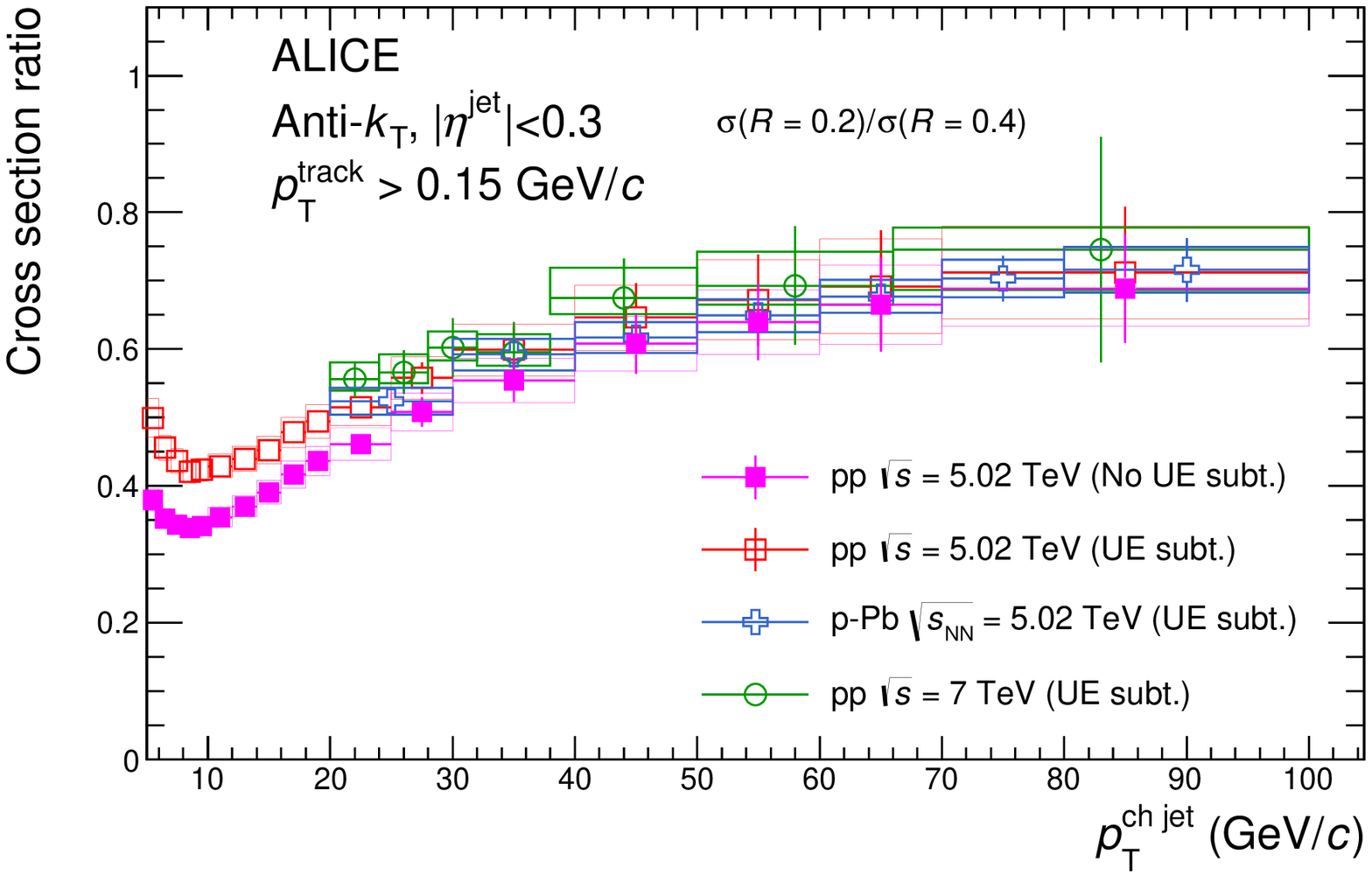}
   \includegraphics[width=0.49\textwidth]{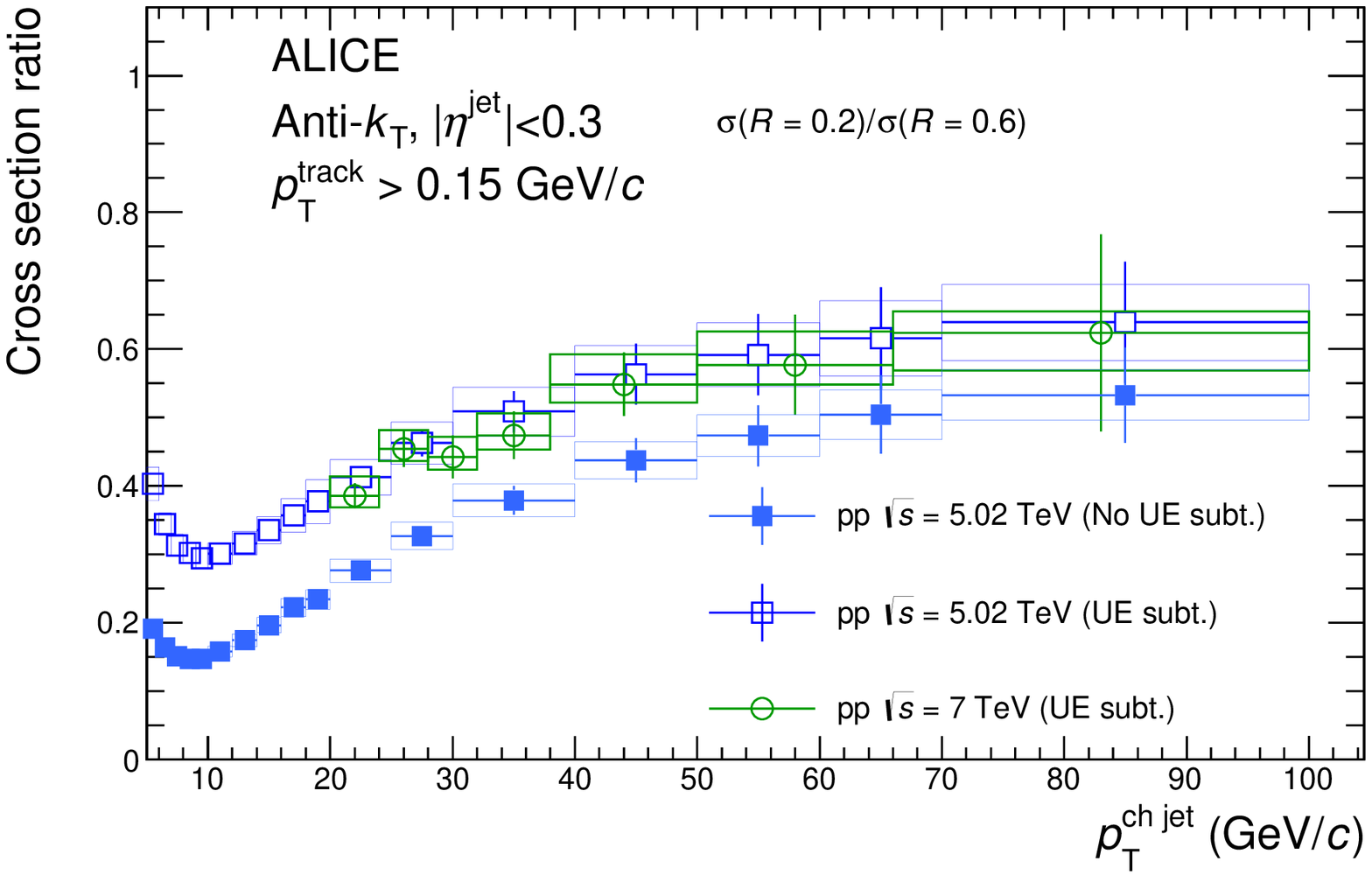}
 \end{center}
 \caption{Charged jet cross section ratio in pp collisions at $\sqrt{s} = 5.02\ \mathrm{TeV}$ is compared to p-Pb collisions at $\sqrt{s_\mathrm{NN}} = 5.02 \ \mathrm{TeV}$~\cite{Adam:2015hoa} and pp collisions at $\sqrt{s} = 7\ \mathrm{TeV}$~\cite{ALICE_chJets7TeV}. Left:  $\sigma(R=0.2)/\sigma(R=0.4)$; Right: $\sigma(R=0.2)/\sigma(R=0.6)$}
 \label{Fig:XsecRatioR0204CompData}
\end{figure*}

Figure~\ref{Fig:XsecRatioR0204CompData} displays a comparison of the results obtained by the ALICE Collaboration in pp collisions at $\sqrt{s} = 7\ \mathrm{TeV}$~\cite{ALICE_chJets7TeV} and p-Pb collisions at $\sqrt{s_\mathrm{NN}} = 5.02 \ \mathrm{TeV}$~\cite{Adam:2015hoa}. All data show a similar increase of the ratio expected from the stronger collimation of jets at higher transverse momentum and agree well within uncertainties. No significant energy dependence nor change with collision species is observed for smaller radii. It should be noted, however, that the earlier ALICE measurements of cross section ratio used for comparison performed the UE subtraction. Since the UE contribution is more pronounced for larger radii ($R = 0.6$), the cross section ratio $\sigma(R=0.2)/\sigma(R=0.6)$ is higher after UE subtraction, and the UE subtracted cross section ratio is consistent to earlier measurements in pp collisions at $\sqrt{s} = 7 \ \mathrm{TeV}$ as presented in Fig. ~\ref{Fig:XsecRatioR0204CompData} (right).
%

%% file: conclusion.tex
The inclusive charged jet cross sections with transverse momentum from $5\ \mathrm{GeV}/c$ to
$100\ \mathrm{GeV}/c$ in pp collisions at $\sqrt{s} = 5.02\ \mathrm{TeV}$ have been measured.
The measurements have been performed using anti-$k_\mathrm{T}$ jet finder algorithm with different 
jet resolution parameters $R = 0.2$, $0.3$, $0.4$, and $0.6$ at mid-rapidity. 
The differential charged jet cross sections are compared with those in LO and NLO pQCD
calculations. There is better agreement between data and predictions at NLO, i.e.\ POWHEG
for parton shower with hadronization by PYTHIA8. The cross section ratios for 
different resolution parameters are also measured, which increase from low to high $p_\mathrm{T}$, and saturate 
at high $p_\mathrm{T}$, indicating that the jet collimation is larger at high $p_\mathrm{T}$.
The ratio for $\sigma(R=0.2) / \sigma(R=0.4)$ is larger than that for $\sigma(R=0.2) / \sigma(R=0.6)$,
and these ratios are consistent with both LO and NLO pQCD calculations.

The data presented in this paper provide an important reference for jet production in QCD, for example the fragmentation function and parton 
distribution functions. It also provides a baseline for the nuclear modification 
factor measurement in Pb-Pb collisions at the same beam energy, 
in order to elucidate the nature of the hot and dense matter created in heavy-ion collisions at the LHC. 
In particular, the results presented in this paper extend the jet measurements to very low $p_\mathrm{T}$, which is challenging to measure in the heavy-ion 
environment due to the UE influence.

%% file: fa_2019-02-11.tex

The ALICE Collaboration would like to thank all its engineers and technicians for their invaluable contributions to the construction of the experiment and the CERN accelerator teams for the outstanding performance of the LHC complex.
The ALICE Collaboration gratefully acknowledges the resources and support provided by all Grid centres and the Worldwide LHC Computing Grid (WLCG) collaboration.
The ALICE Collaboration acknowledges the following funding agencies for their support in building and running the ALICE detector:
A. I. Alikhanyan National Science Laboratory (Yerevan Physics Institute) Foundation (ANSL), State Committee of Science and World Federation of Scientists (WFS), Armenia;
Austrian Academy of Sciences, Austrian Science Fund (FWF): [M 2467-N36] and Nationalstiftung f\"{u}r Forschung, Technologie und Entwicklung, Austria;
Ministry of Communications and High Technologies, National Nuclear Research Center, Azerbaijan;
Conselho Nacional de Desenvolvimento Cient\'{\i}fico e Tecnol\'{o}gico (CNPq), Universidade Federal do Rio Grande do Sul (UFRGS), Financiadora de Estudos e Projetos (Finep) and Funda\c{c}\~{a}o de Amparo \`{a} Pesquisa do Estado de S\~{a}o Paulo (FAPESP), Brazil;
Ministry of Science \& Technology of China (MSTC), National Natural Science Foundation of China (NSFC) and Ministry of Education of China (MOEC) , China;
Croatian Science Foundation and Ministry of Science and Education, Croatia;
Centro de Aplicaciones Tecnol\'{o}gicas y Desarrollo Nuclear (CEADEN), Cubaenerg\'{\i}a, Cuba;
Ministry of Education, Youth and Sports of the Czech Republic, Czech Republic;
The Danish Council for Independent Research | Natural Sciences, the Carlsberg Foundation and Danish National Research Foundation (DNRF), Denmark;
Helsinki Institute of Physics (HIP), Finland;
Commissariat \`{a} l'Energie Atomique (CEA), Institut National de Physique Nucl\'{e}aire et de Physique des Particules (IN2P3) and Centre National de la Recherche Scientifique (CNRS) and R\'{e}gion des  Pays de la Loire, France;
Bundesministerium f\"{u}r Bildung und Forschung (BMBF) and GSI Helmholtzzentrum f\"{u}r Schwerionenforschung GmbH, Germany;
General Secretariat for Research and Technology, Ministry of Education, Research and Religions, Greece;
National Research, Development and Innovation Office, Hungary;
Department of Atomic Energy Government of India (DAE), Department of Science and Technology, Government of India (DST), University Grants Commission, Government of India (UGC) and Council of Scientific and Industrial Research (CSIR), India;
Indonesian Institute of Science, Indonesia;
Centro Fermi - Museo Storico della Fisica e Centro Studi e Ricerche Enrico Fermi and Istituto Nazionale di Fisica Nucleare (INFN), Italy;
Institute for Innovative Science and Technology , Nagasaki Institute of Applied Science (IIST), Japan Society for the Promotion of Science (JSPS) KAKENHI and Japanese Ministry of Education, Culture, Sports, Science and Technology (MEXT), Japan;
Consejo Nacional de Ciencia (CONACYT) y Tecnolog\'{i}a, through Fondo de Cooperaci\'{o}n Internacional en Ciencia y Tecnolog\'{i}a (FONCICYT) and Direcci\'{o}n General de Asuntos del Personal Academico (DGAPA), Mexico;
Nederlandse Organisatie voor Wetenschappelijk Onderzoek (NWO), Netherlands;
The Research Council of Norway, Norway;
Commission on Science and Technology for Sustainable Development in the South (COMSATS), Pakistan;
Pontificia Universidad Cat\'{o}lica del Per\'{u}, Peru;
Ministry of Science and Higher Education and National Science Centre, Poland;
Korea Institute of Science and Technology Information and National Research Foundation of Korea (NRF), Republic of Korea;
Ministry of Education and Scientific Research, Institute of Atomic Physics and Ministry of Research and Innovation and Institute of Atomic Physics, Romania;
Joint Institute for Nuclear Research (JINR), Ministry of Education and Science of the Russian Federation, National Research Centre Kurchatov Institute, Russian Science Foundation and Russian Foundation for Basic Research, Russia;
Ministry of Education, Science, Research and Sport of the Slovak Republic, Slovakia;
National Research Foundation of South Africa, South Africa;
Swedish Research Council (VR) and Knut \& Alice Wallenberg Foundation (KAW), Sweden;
European Organization for Nuclear Research, Switzerland;
National Science and Technology Development Agency (NSDTA), Suranaree University of Technology (SUT) and Office of the Higher Education Commission under NRU project of Thailand, Thailand;
Turkish Atomic Energy Agency (TAEK), Turkey;
National Academy of  Sciences of Ukraine, Ukraine;
Science and Technology Facilities Council (STFC), United Kingdom;
National Science Foundation of the United States of America (NSF) and United States Department of Energy, Office of Nuclear Physics (DOE NP), United States of America.

%% file: 2019-02-11-Alice_Authorlist_2019-Feb-11.tex

\begingroup
\small
\begin{flushleft}
S.~Acharya\Irefn{org141}\And 
D.~Adamov\'{a}\Irefn{org93}\And 
S.P.~Adhya\Irefn{org141}\And 
A.~Adler\Irefn{org74}\And 
J.~Adolfsson\Irefn{org80}\And 
M.M.~Aggarwal\Irefn{org98}\And 
G.~Aglieri Rinella\Irefn{org34}\And 
M.~Agnello\Irefn{org31}\And 
N.~Agrawal\Irefn{org10}\And 
Z.~Ahammed\Irefn{org141}
% ...
\And
\renewcommand\labelenumi{\textsuperscript{\theenumi}~}

\section*{Affiliation notes}
\renewcommand\theenumi{\roman{enumi}}
\begin{Authlist}
\item \Adef{org*}Deceased
\item \Adef{orgI}Dipartimento DET del Politecnico di Torino, Turin, Italy
\item \Adef{orgII}M.V. Lomonosov Moscow State University, D.V. Skobeltsyn Institute of Nuclear, Physics, Moscow, Russia
\item \Adef{orgIII}Department of Applied Physics, Aligarh Muslim University, Aligarh, India
\item \Adef{orgIV}Institute of Theoretical Physics, University of Wroclaw, Poland
\end{Authlist}

\section*{Collaboration Institutes}
\renewcommand\theenumi{\arabic{enumi}~}
\begin{Authlist}
\item \Idef{org1}A.I. Alikhanyan National Science Laboratory (Yerevan Physics Institute) Foundation, Yerevan, Armenia
\item \Idef{org2}Bogolyubov Institute for Theoretical Physics, National Academy of Sciences of Ukraine, Kiev, Ukraine
\item \Idef{org3}Bose Institute, Department of Physics  and Centre for Astroparticle Physics and Space Science (CAPSS), Kolkata, India
\item \Idef{org4}Budker Institute for Nuclear Physics, Novosibirsk, Russia
\item \Idef{org5}California Polytechnic State University, San Luis Obispo, California, United States
\item \Idef{org6}Central China Normal University, Wuhan, China
\item \Idef{org7}Centre de Calcul de l'IN2P3, Villeurbanne, Lyon, France
\item \Idef{org8}Centro de Aplicaciones Tecnol\'{o}gicas y Desarrollo Nuclear (CEADEN), Havana, Cuba
\item \Idef{org9}Centro de Investigaci\'{o}n y de Estudios Avanzados (CINVESTAV), Mexico City and M\'{e}rida, Mexico
\item \Idef{org10}Centro Fermi - Museo Storico della Fisica e Centro Studi e Ricerche ``Enrico Fermi', Rome, Italy
\item \Idef{org11}Chicago State University, Chicago, Illinois, United States
\item \Idef{org12}China Institute of Atomic Energy, Beijing, China
\item \Idef{org13}Chonbuk National University, Jeonju, Republic of Korea
\item \Idef{org14}Comenius University Bratislava, Faculty of Mathematics, Physics and Informatics, Bratislava, Slovakia
\item \Idef{org15}COMSATS University Islamabad, Islamabad, Pakistan
\item \Idef{org16}Creighton University, Omaha, Nebraska, United States
\item \Idef{org17}Department of Physics, Aligarh Muslim University, Aligarh, India
\item \Idef{org18}Department of Physics, Pusan National University, Pusan, Republic of Korea
\item \Idef{org19}Department of Physics, Sejong University, Seoul, Republic of Korea
\item \Idef{org20}Department of Physics, University of California, Berkeley, California, United States
\item \Idef{org21}Department of Physics, University of Oslo, Oslo, Norway
\item \Idef{org22}Department of Physics and Technology, University of Bergen, Bergen, Norway
\item \Idef{org23}Dipartimento di Fisica dell'Universit\`{a} 'La Sapienza' and Sezione INFN, Rome, Italy
\item \Idef{org24}Dipartimento di Fisica dell'Universit\`{a} and Sezione INFN, Cagliari, Italy
\item \Idef{org25}Dipartimento di Fisica dell'Universit\`{a} and Sezione INFN, Trieste, Italy
\item \Idef{org26}Dipartimento di Fisica dell'Universit\`{a} and Sezione INFN, Turin, Italy
\item \Idef{org27}Dipartimento di Fisica e Astronomia dell'Universit\`{a} and Sezione INFN, Bologna, Italy
\item \Idef{org28}Dipartimento di Fisica e Astronomia dell'Universit\`{a} and Sezione INFN, Catania, Italy
\item \Idef{org29}Dipartimento di Fisica e Astronomia dell'Universit\`{a} and Sezione INFN, Padova, Italy
\item \Idef{org30}Dipartimento di Fisica `E.R.~Caianiello' dell'Universit\`{a} and Gruppo Collegato INFN, Salerno, Italy
\item \Idef{org31}Dipartimento DISAT del Politecnico and Sezione INFN, Turin, Italy
\item \Idef{org32}Dipartimento di Scienze e Innovazione Tecnologica dell'Universit\`{a} del Piemonte Orientale and INFN Sezione di Torino, Alessandria, Italy
\item \Idef{org33}Dipartimento Interateneo di Fisica `M.~Merlin' and Sezione INFN, Bari, Italy
\item \Idef{org34}European Organization for Nuclear Research (CERN), Geneva, Switzerland
\item \Idef{org35}Faculty of Electrical Engineering, Mechanical Engineering and Naval Architecture, University of Split, Split, Croatia
\item \Idef{org36}Faculty of Engineering and Science, Western Norway University of Applied Sciences, Bergen, Norway
\item \Idef{org37}Faculty of Nuclear Sciences and Physical Engineering, Czech Technical University in Prague, Prague, Czech Republic
\item \Idef{org38}Faculty of Science, P.J.~\v{S}af\'{a}rik University, Ko\v{s}ice, Slovakia
\item \Idef{org39}Frankfurt Institute for Advanced Studies, Johann Wolfgang Goethe-Universit\"{a}t Frankfurt, Frankfurt, Germany
\item \Idef{org40}Gangneung-Wonju National University, Gangneung, Republic of Korea
\item \Idef{org41}Gauhati University, Department of Physics, Guwahati, India
\item \Idef{org42}Helmholtz-Institut f\"{u}r Strahlen- und Kernphysik, Rheinische Friedrich-Wilhelms-Universit\"{a}t Bonn, Bonn, Germany
\item \Idef{org43}Helsinki Institute of Physics (HIP), Helsinki, Finland
\item \Idef{org44}High Energy Physics Group,  Universidad Aut\'{o}noma de Puebla, Puebla, Mexico
\item \Idef{org45}Hiroshima University, Hiroshima, Japan
\item \Idef{org46}Hochschule Worms, Zentrum  f\"{u}r Technologietransfer und Telekommunikation (ZTT), Worms, Germany
\item \Idef{org47}Horia Hulubei National Institute of Physics and Nuclear Engineering, Bucharest, Romania
\item \Idef{org48}Indian Institute of Technology Bombay (IIT), Mumbai, India
\item \Idef{org49}Indian Institute of Technology Indore, Indore, India
\item \Idef{org50}Indonesian Institute of Sciences, Jakarta, Indonesia
\item \Idef{org51}INFN, Laboratori Nazionali di Frascati, Frascati, Italy
\item \Idef{org52}INFN, Sezione di Bari, Bari, Italy
\item \Idef{org53}INFN, Sezione di Bologna, Bologna, Italy
\item \Idef{org54}INFN, Sezione di Cagliari, Cagliari, Italy
\item \Idef{org55}INFN, Sezione di Catania, Catania, Italy
\item \Idef{org56}INFN, Sezione di Padova, Padova, Italy
\item \Idef{org57}INFN, Sezione di Roma, Rome, Italy
\item \Idef{org58}INFN, Sezione di Torino, Turin, Italy
\item \Idef{org59}INFN, Sezione di Trieste, Trieste, Italy
\item \Idef{org60}Inha University, Incheon, Republic of Korea
\item \Idef{org61}Institut de Physique Nucl\'{e}aire d'Orsay (IPNO), Institut National de Physique Nucl\'{e}aire et de Physique des Particules (IN2P3/CNRS), Universit\'{e} de Paris-Sud, Universit\'{e} Paris-Saclay, Orsay, France
\item \Idef{org62}Institute for Nuclear Research, Academy of Sciences, Moscow, Russia
\item \Idef{org63}Institute for Subatomic Physics, Utrecht University/Nikhef, Utrecht, Netherlands
\item \Idef{org64}Institute for Theoretical and Experimental Physics, Moscow, Russia
\item \Idef{org65}Institute of Experimental Physics, Slovak Academy of Sciences, Ko\v{s}ice, Slovakia
\item \Idef{org66}Institute of Physics, Homi Bhabha National Institute, Bhubaneswar, India
\item \Idef{org67}Institute of Physics of the Czech Academy of Sciences, Prague, Czech Republic
\item \Idef{org68}Institute of Space Science (ISS), Bucharest, Romania
\item \Idef{org69}Institut f\"{u}r Kernphysik, Johann Wolfgang Goethe-Universit\"{a}t Frankfurt, Frankfurt, Germany
\item \Idef{org70}Instituto de Ciencias Nucleares, Universidad Nacional Aut\'{o}noma de M\'{e}xico, Mexico City, Mexico
\item \Idef{org71}Instituto de F\'{i}sica, Universidade Federal do Rio Grande do Sul (UFRGS), Porto Alegre, Brazil
\item \Idef{org72}Instituto de F\'{\i}sica, Universidad Nacional Aut\'{o}noma de M\'{e}xico, Mexico City, Mexico
\item \Idef{org73}iThemba LABS, National Research Foundation, Somerset West, South Africa
\item \Idef{org74}Johann-Wolfgang-Goethe Universit\"{a}t Frankfurt Institut f\"{u}r Informatik, Fachbereich Informatik und Mathematik, Frankfurt, Germany
\item \Idef{org75}Joint Institute for Nuclear Research (JINR), Dubna, Russia
\item \Idef{org76}Korea Institute of Science and Technology Information, Daejeon, Republic of Korea
\item \Idef{org77}KTO Karatay University, Konya, Turkey
\item \Idef{org78}Laboratoire de Physique Subatomique et de Cosmologie, Universit\'{e} Grenoble-Alpes, CNRS-IN2P3, Grenoble, France
\item \Idef{org79}Lawrence Berkeley National Laboratory, Berkeley, California, United States
\item \Idef{org80}Lund University Department of Physics, Division of Particle Physics, Lund, Sweden
\item \Idef{org81}Nagasaki Institute of Applied Science, Nagasaki, Japan
\item \Idef{org82}Nara Women{'}s University (NWU), Nara, Japan
\item \Idef{org83}National and Kapodistrian University of Athens, School of Science, Department of Physics , Athens, Greece
\item \Idef{org84}National Centre for Nuclear Research, Warsaw, Poland
\item \Idef{org85}National Institute of Science Education and Research, Homi Bhabha National Institute, Jatni, India
\item \Idef{org86}National Nuclear Research Center, Baku, Azerbaijan
\item \Idef{org87}National Research Centre Kurchatov Institute, Moscow, Russia
\item \Idef{org88}Niels Bohr Institute, University of Copenhagen, Copenhagen, Denmark
\item \Idef{org89}Nikhef, National institute for subatomic physics, Amsterdam, Netherlands
\item \Idef{org90}NRC Kurchatov Institute IHEP, Protvino, Russia
\item \Idef{org91}NRNU Moscow Engineering Physics Institute, Moscow, Russia
\item \Idef{org92}Nuclear Physics Group, STFC Daresbury Laboratory, Daresbury, United Kingdom
\item \Idef{org93}Nuclear Physics Institute of the Czech Academy of Sciences, \v{R}e\v{z} u Prahy, Czech Republic
\item \Idef{org94}Oak Ridge National Laboratory, Oak Ridge, Tennessee, United States
\item \Idef{org95}Ohio State University, Columbus, Ohio, United States
\item \Idef{org96}Petersburg Nuclear Physics Institute, Gatchina, Russia
\item \Idef{org97}Physics department, Faculty of science, University of Zagreb, Zagreb, Croatia
\item \Idef{org98}Physics Department, Panjab University, Chandigarh, India
\item \Idef{org99}Physics Department, University of Jammu, Jammu, India
\item \Idef{org100}Physics Department, University of Rajasthan, Jaipur, India
\item \Idef{org101}Physikalisches Institut, Eberhard-Karls-Universit\"{a}t T\"{u}bingen, T\"{u}bingen, Germany
\item \Idef{org102}Physikalisches Institut, Ruprecht-Karls-Universit\"{a}t Heidelberg, Heidelberg, Germany
\item \Idef{org103}Physik Department, Technische Universit\"{a}t M\"{u}nchen, Munich, Germany
\item \Idef{org104}Politecnico di Bari, Bari, Italy
\item \Idef{org105}Research Division and ExtreMe Matter Institute EMMI, GSI Helmholtzzentrum f\"ur Schwerionenforschung GmbH, Darmstadt, Germany
\item \Idef{org106}Rudjer Bo\v{s}kovi\'{c} Institute, Zagreb, Croatia
\item \Idef{org107}Russian Federal Nuclear Center (VNIIEF), Sarov, Russia
\item \Idef{org108}Saha Institute of Nuclear Physics, Homi Bhabha National Institute, Kolkata, India
\item \Idef{org109}School of Physics and Astronomy, University of Birmingham, Birmingham, United Kingdom
\item \Idef{org110}Secci\'{o}n F\'{\i}sica, Departamento de Ciencias, Pontificia Universidad Cat\'{o}lica del Per\'{u}, Lima, Peru
\item \Idef{org111}Shanghai Institute of Applied Physics, Shanghai, China
\item \Idef{org112}St. Petersburg State University, St. Petersburg, Russia
\item \Idef{org113}Stefan Meyer Institut f\"{u}r Subatomare Physik (SMI), Vienna, Austria
\item \Idef{org114}SUBATECH, IMT Atlantique, Universit\'{e} de Nantes, CNRS-IN2P3, Nantes, France
\item \Idef{org115}Suranaree University of Technology, Nakhon Ratchasima, Thailand
\item \Idef{org116}Technical University of Ko\v{s}ice, Ko\v{s}ice, Slovakia
\item \Idef{org117}Technische Universit\"{a}t M\"{u}nchen, Excellence Cluster 'Universe', Munich, Germany
\item \Idef{org118}The Henryk Niewodniczanski Institute of Nuclear Physics, Polish Academy of Sciences, Cracow, Poland
\item \Idef{org119}The University of Texas at Austin, Austin, Texas, United States
\item \Idef{org120}Universidad Aut\'{o}noma de Sinaloa, Culiac\'{a}n, Mexico
\item \Idef{org121}Universidade de S\~{a}o Paulo (USP), S\~{a}o Paulo, Brazil
\item \Idef{org122}Universidade Estadual de Campinas (UNICAMP), Campinas, Brazil
\item \Idef{org123}Universidade Federal do ABC, Santo Andre, Brazil
\item \Idef{org124}University College of Southeast Norway, Tonsberg, Norway
\item \Idef{org125}University of Cape Town, Cape Town, South Africa
\item \Idef{org126}University of Houston, Houston, Texas, United States
\item \Idef{org127}University of Jyv\"{a}skyl\"{a}, Jyv\"{a}skyl\"{a}, Finland
\item \Idef{org128}University of Liverpool, Liverpool, United Kingdom
\item \Idef{org129}University of Science and Techonology of China, Hefei, China
\item \Idef{org130}University of Tennessee, Knoxville, Tennessee, United States
\item \Idef{org131}University of the Witwatersrand, Johannesburg, South Africa
\item \Idef{org132}University of Tokyo, Tokyo, Japan
\item \Idef{org133}University of Tsukuba, Tsukuba, Japan
\item \Idef{org134}Universit\'{e} Clermont Auvergne, CNRS/IN2P3, LPC, Clermont-Ferrand, France
\item \Idef{org135}Universit\'{e} de Lyon, Universit\'{e} Lyon 1, CNRS/IN2P3, IPN-Lyon, Villeurbanne, Lyon, France
\item \Idef{org136}Universit\'{e} de Strasbourg, CNRS, IPHC UMR 7178, F-67000 Strasbourg, France, Strasbourg, France
\item \Idef{org137}Universit\'{e} Paris-Saclay Centre d'Etudes de Saclay (CEA), IRFU, D\'{e}partment de Physique Nucl\'{e}aire (DPhN), Saclay, France
\item \Idef{org138}Universit\`{a} degli Studi di Foggia, Foggia, Italy
\item \Idef{org139}Universit\`{a} degli Studi di Pavia, Pavia, Italy
\item \Idef{org140}Universit\`{a} di Brescia, Brescia, Italy
\item \Idef{org141}Variable Energy Cyclotron Centre, Homi Bhabha National Institute, Kolkata, India
\item \Idef{org142}Warsaw University of Technology, Warsaw, Poland
\item \Idef{org143}Wayne State University, Detroit, Michigan, United States
\item \Idef{org144}Westf\"{a}lische Wilhelms-Universit\"{a}t M\"{u}nster, Institut f\"{u}r Kernphysik, M\"{u}nster, Germany
\item \Idef{org145}Wigner Research Centre for Physics, Hungarian Academy of Sciences, Budapest, Hungary
\item \Idef{org146}Yale University, New Haven, Connecticut, United States
\item \Idef{org147}Yonsei University, Seoul, Republic of Korea
\end{Authlist}
\endgroup

%% file: appendix.tex
\subsection{Charged jet cross section and ratios}
\begin{figure*}[htbp]
 \begin{center}
 \includegraphics[width=0.8\textwidth]{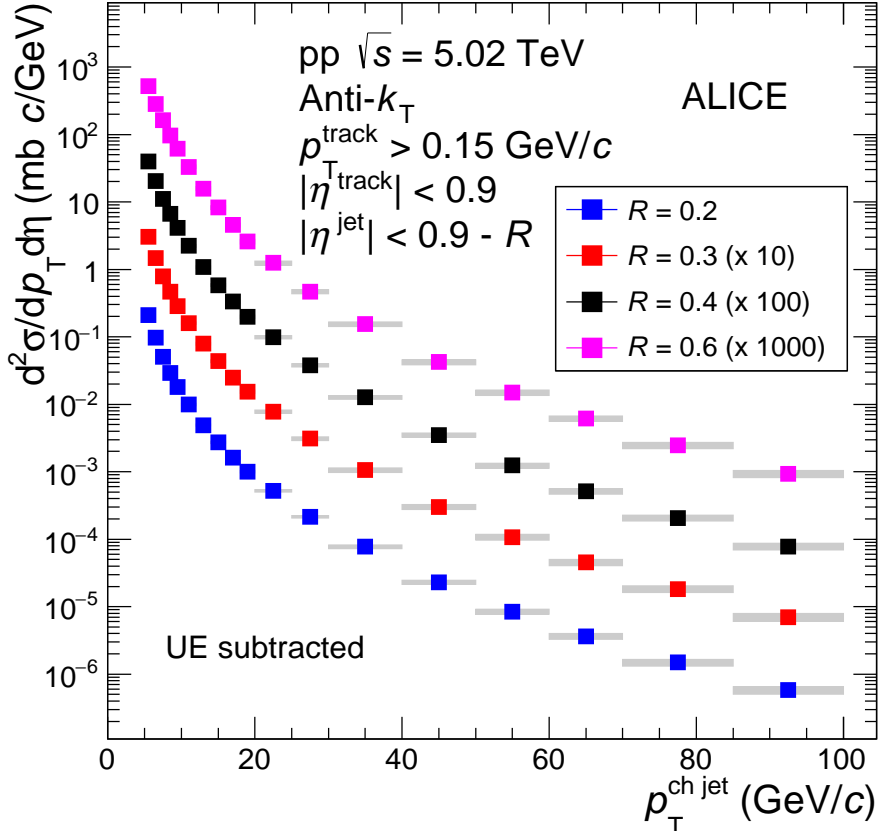}
 \end{center}
 \caption{Charged jet differential cross sections after the detector effects correction and UE subtraction in pp collisions at $\sqrt{s} = 5.02\ \mathrm{TeV}$. Statistical uncertainties are displayed as vertical error bars. The total systematic uncertainties are shown as shaded bands around the data points. Data are scaled to enhance visibility.}
 \label{Fig:Xsec_UE}
\end{figure*}

\begin{figure*}[htbp]
 \begin{center}
  \includegraphics[width=0.85\textwidth]{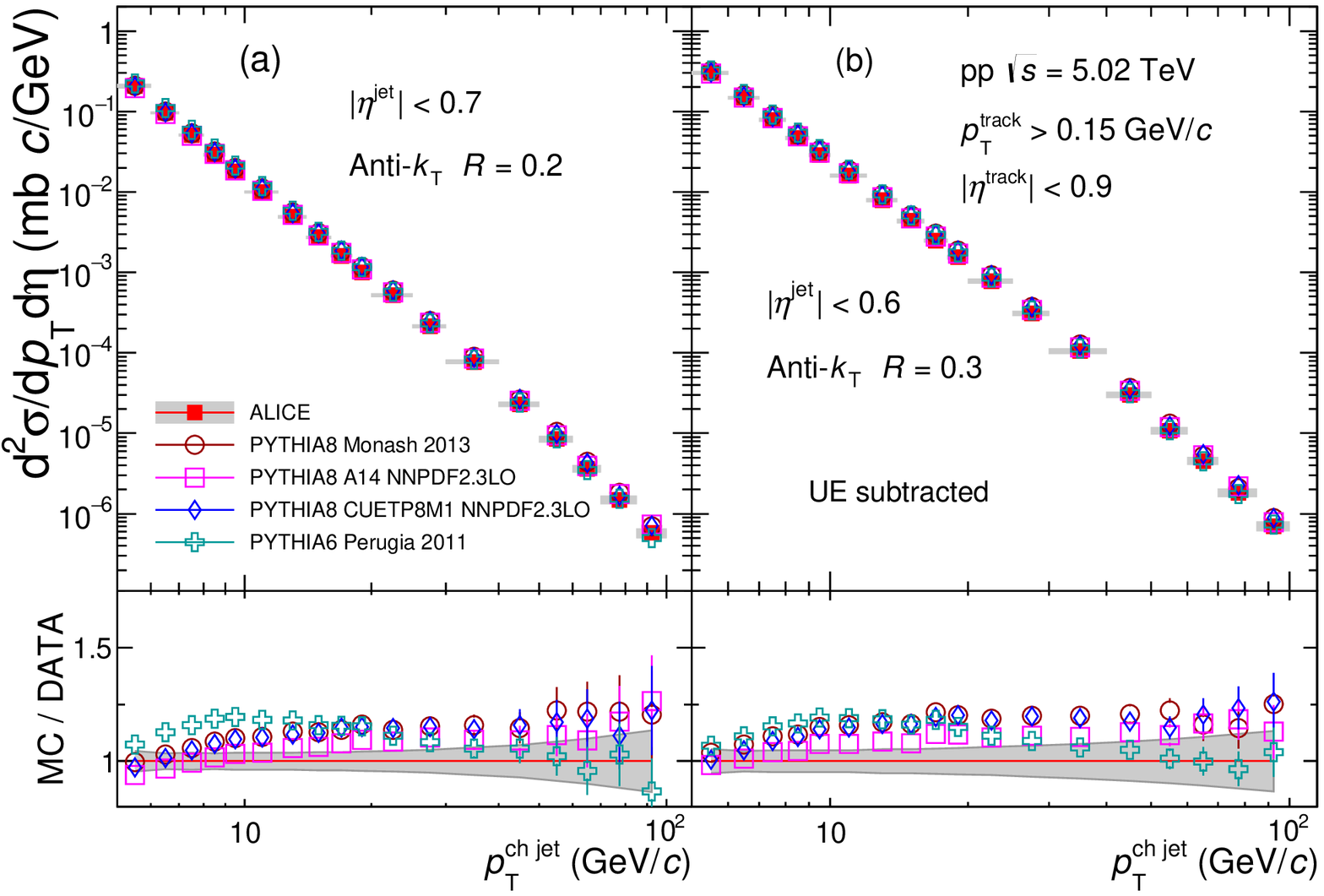} \\  
  \includegraphics[width=0.85\textwidth]{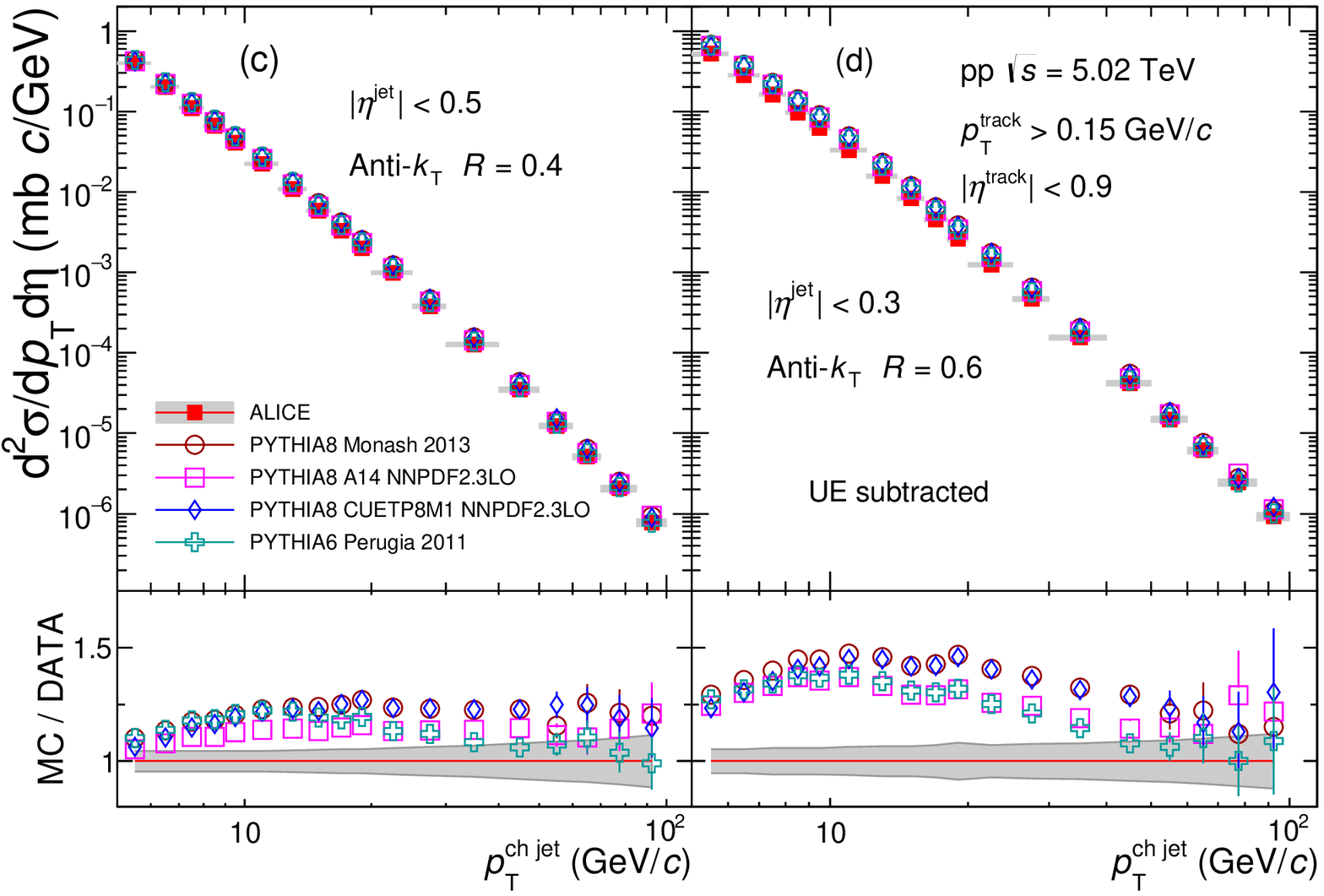}  
 \end{center}
 \caption{Comparison of the charged jet cross section to LO MC predictions with different 
jet resolution parameter $R=0.2$ (a), $0.3$ (b), $0.4$ (c), and $0.6$ (d). UE subtraction is applied. Statistical uncertainties are displayed as vertical error bars. The systematic uncertainty on the data is indicated by a shaded band drawn around unity. The red lines in the ratio correspond to unity.}
 \label{Fig:XsecCompMCUESub}
\end{figure*}

\begin{figure*}[htbp]
 \begin{center}
  \includegraphics[width=0.85\textwidth]{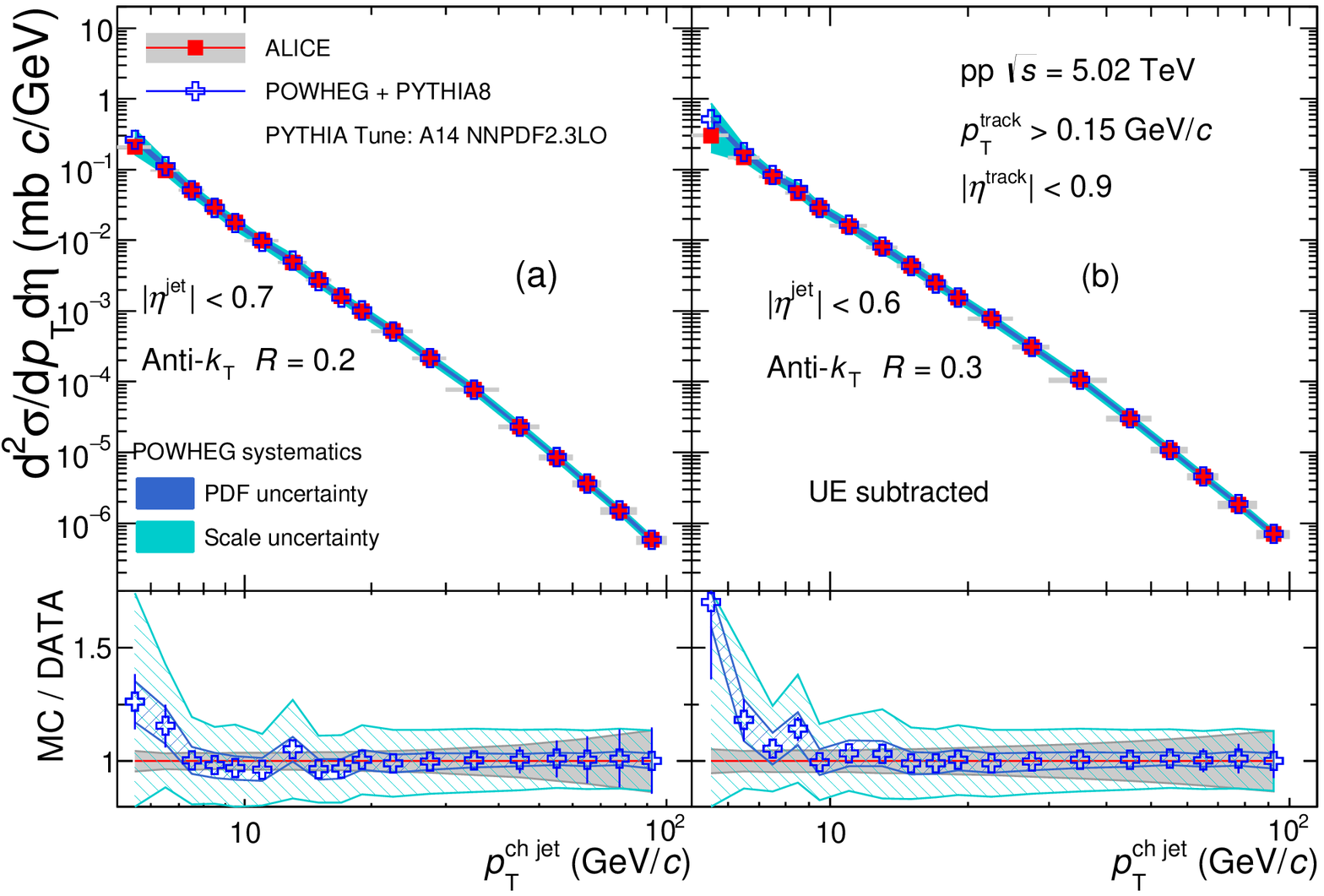} \\  
  \includegraphics[width=0.85\textwidth]{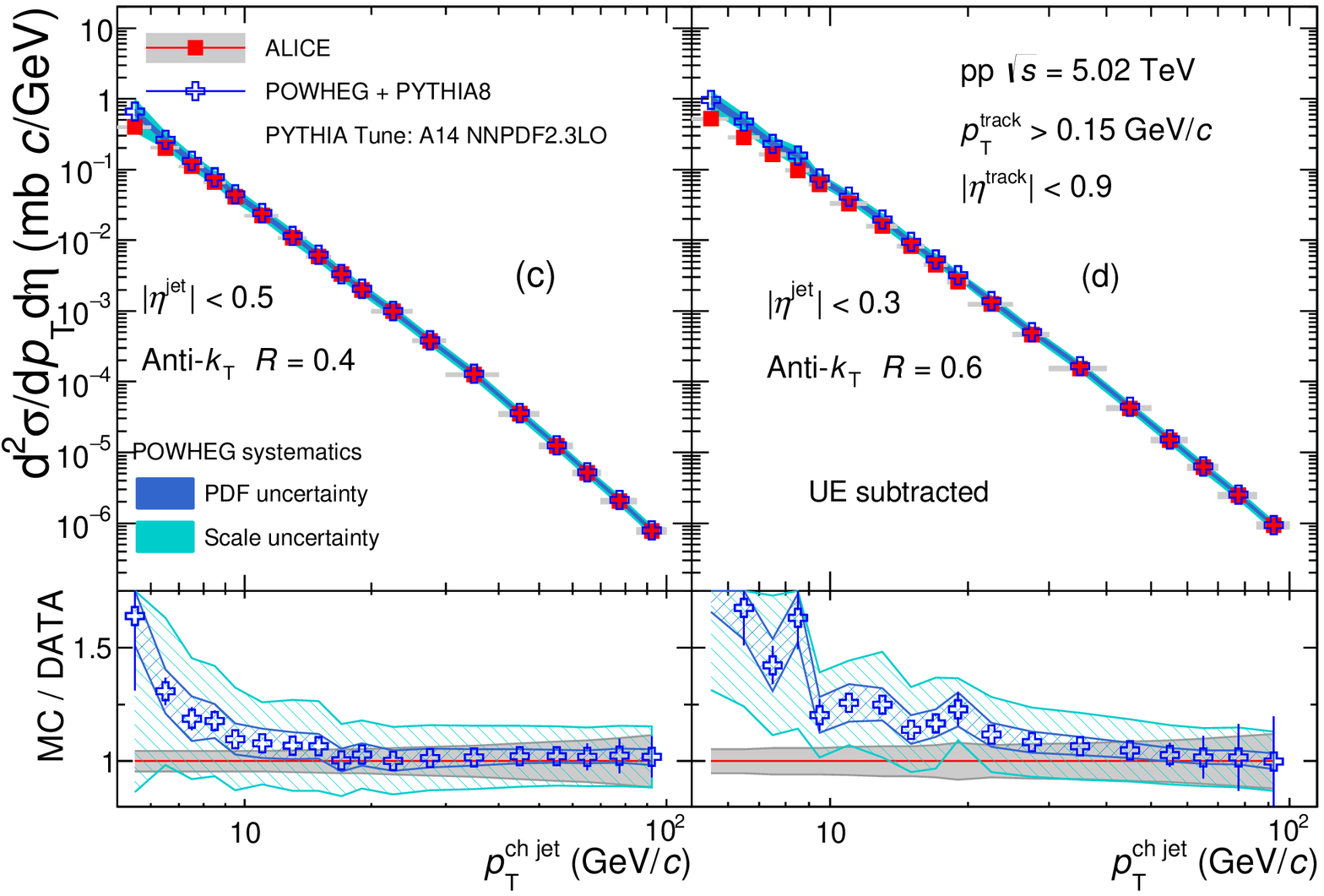}  
 \end{center}
 \caption{Comparison of the charged jet cross section to NLO MC prediction (POWHEG+PYTHIA8) with different 
jet resolution parameter $R=0.2$ (a), $0.3$ (b), $0.4$ (c), and $0.6$ (d). UE subtraction is applied. Statistical uncertainties are displayed as vertical error bars. The systematic uncertainty on the data is indicated by a shaded band drawn around unity. The red lines in the ratio correspond to unity.}
 \label{Fig:XsecCompMCNLOUESub}
\end{figure*}

The inclusive charged jet cross sections after the detector effects correction and UE subtraction using the anti-$k_{T}$ jet finder in pp collisions at $\sqrt{s} = 5.02\ \mathrm{TeV}$ are presented in Fig.~\ref{Fig:Xsec_UE},  The comparisons to different LO and NLO theoretical calculations are shown in Fig.~\ref{Fig:XsecCompMCUESub} and Fig.~\ref{Fig:XsecCompMCNLOUESub}, respectively. The UE contamination is corrected on an event-by-event basis by the perpendicular cone estimator.

\begin{figure*}[htbp]
 \begin{center}
   \includegraphics[width=0.8\textwidth]{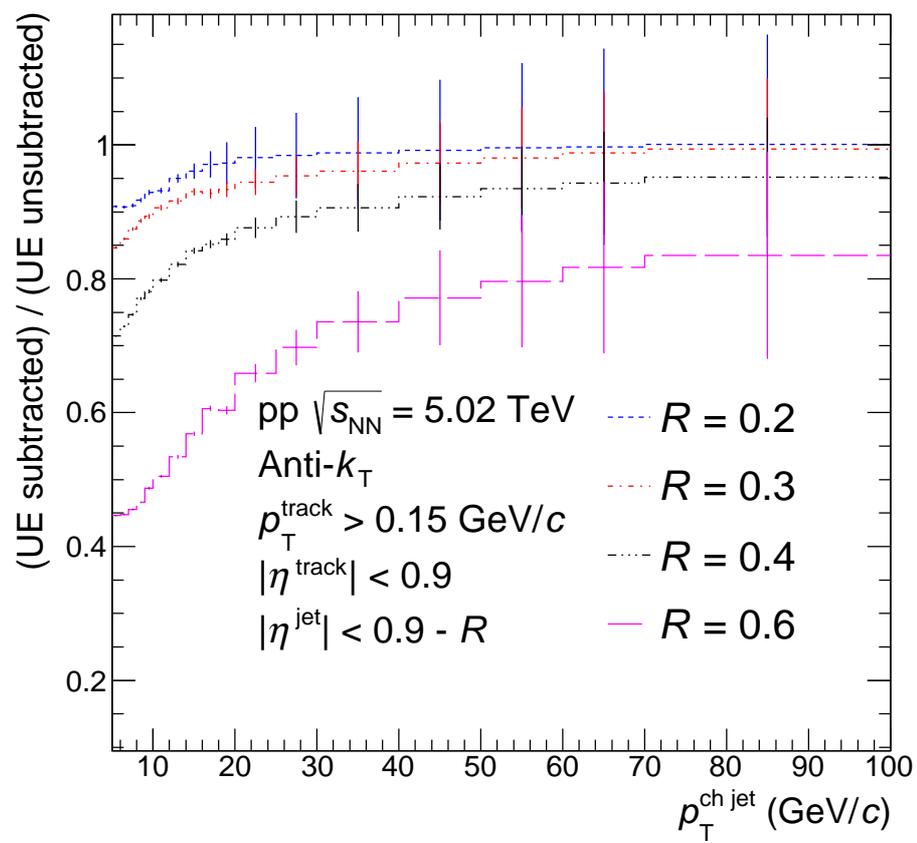}
 \end{center}
 \caption{Charged jet differential cross section ratio with and without UE subtraction in pp collisions at $\sqrt{s} = 5.02\ \mathrm{TeV}$.}
 \label{Fig:UERatio}
\end{figure*}

The impact of the UE subtraction on the inclusive jet spectrum can be seen in Fig.~\ref{Fig:UERatio}, which is the jet cross section ratio with (Fig.~\ref{Fig:Xsec_UE}) and without UE (Fig.~\ref{Fig:Xsec}) subtraction.  After the UE subtraction, the agreement between data and MC becomes worse, since current MC tunes do not model the UE production mechanism in proton-proton collisions well.

\newpage

\subsection{Jet cross sections with leading track cut}
\begin{figure*}[htbp]
 \begin{center}
   \includegraphics[width=0.8\textwidth]{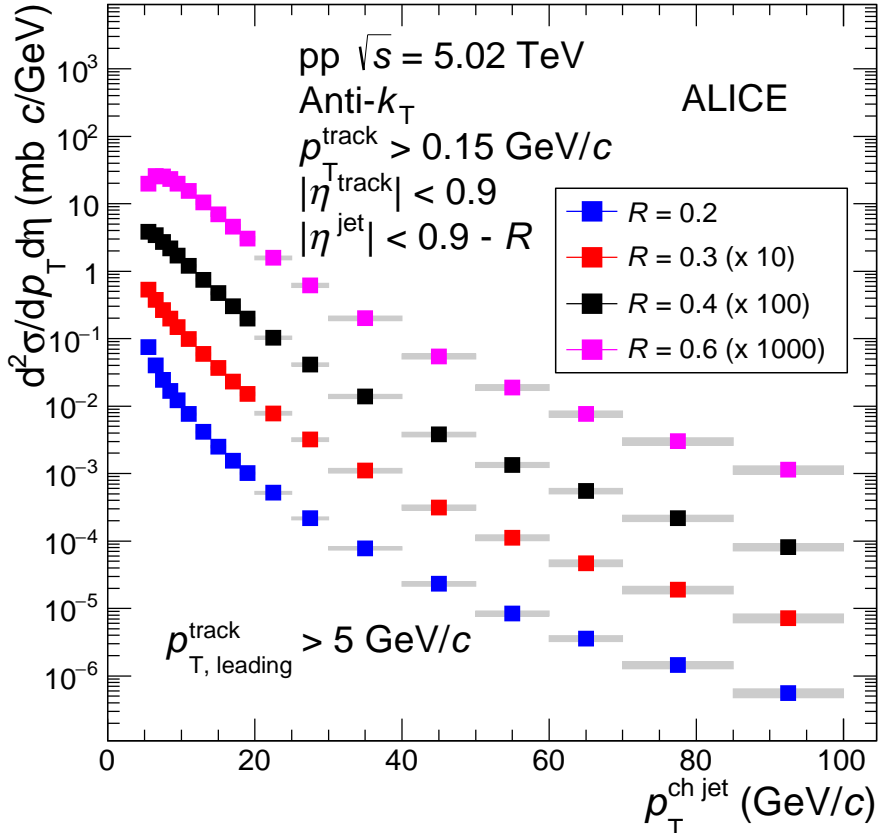}
 \end{center}
 \caption{Charged jet differential cross sections without UE subtraction in pp collisions at $\sqrt{s} = 5.02\ \mathrm{TeV}$ with the leading track bias. All jets must contain at least one track with $p_\mathrm{T} > 5\ \mathrm{GeV}/c$.
Statistical uncertainties are displayed as vertical error bars. The total systematic uncertainties are shown as shaded bands around the data points. Data are scaled to enhance visibility.}
 \label{Fig:XsecBiased}
\end{figure*}

\begin{figure*}[htbp]
 \begin{center}
   \includegraphics[width=1.\textwidth]{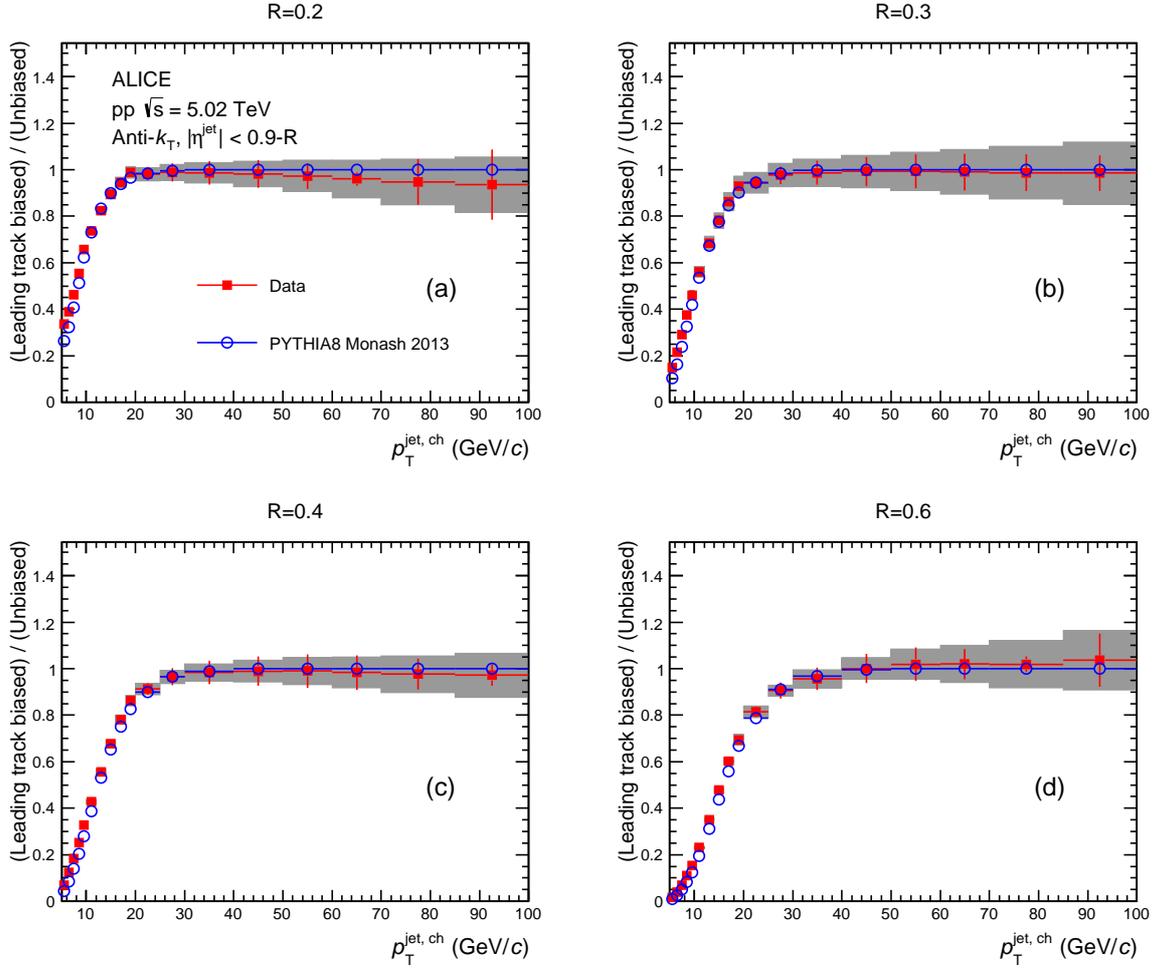}
 \end{center}
 \caption{Charged jet differential cross section ratios with and without leading track bias in pp collisions at $\sqrt{s} = 5.02\ \mathrm{TeV}$ with different 
jet resolution parameters $R=0.2$ (top left), $0.3$ (top right), $0.4$ (bottom left), and $0.6$ (bottom right). Without UE subtraction.  For biased results, all jets must contain at least one track with $p_\mathrm{T} > 5\ \mathrm{GeV}/c$.
Statistical uncertainties are displayed as vertical error bars. The systematic uncertainties are shown as shaded bands around the data points.}
 \label{Fig:XsecRatioBiased}
\end{figure*}

The fully corrected inclusive charged jet cross sections by requiring at least one track with $p_\mathrm{T} > 5\ \mathrm{GeV}/c$ using the anti-$k_{T}$ jet finder in pp collisions at $\sqrt{s} = 5.02\ \mathrm{TeV}$ are presented in Fig.~\ref{Fig:XsecBiased}. The ratio of the cross section with and without leading track cut bias is shown in Fig.~\ref{Fig:XsecRatioBiased}. The jet cross sections are without UE subtraction in this section.

\label{appendixB}

